\documentclass[traditabstract]{aa}  
\usepackage{newtxmath}
\usepackage{newtxtext}
\usepackage[T1]{fontenc}

\DeclareRobustCommand{\VAN}[3]{#2}
\let\VANthebibliography\thebibliography
\def\thebibliography{\DeclareRobustCommand{\VAN}[3]{##3}\VANthebibliography}

\usepackage{graphicx}	% Including figure files
\usepackage{amsmath}	% Advanced maths commands
\usepackage[dvipsnames,usenames]{color} 
\usepackage{float} 
\usepackage{natbib}
\usepackage{adjustbox}
\usepackage[normalem]{ulem}
\usepackage{enumitem}
\usepackage{caption}
\usepackage{booktabs}
\usepackage{placeins}

\newcommand{\rmd}{{\rm d}}

\newcommand{\ie}{{\it i.e.~}}
\newcommand{\cf}{{\it cf.~}}
\newcommand{\eg}{{\it e.g.~}}
\newcommand{\iec}{{\it i.e.,~}}

\newcommand{\egc}{{\it e.g.,~}}
\newcommand{\dpias}{{$\Delta\Pi_{\rm as}$}}
\newcommand{\wgstart}{{$\tilde\omega_{\rm g}^\star$}}
\newcommand{\wgstar}{{$\omega_{\rm g}^\star$}}
\newcommand{\phase}{{$\delta$}}
\newcommand{\as}{{$A_{\rm st}$}}
\newcommand{\ag}{{$A_{\rm G}$}}
\newcommand{\width}{{$\Delta_{\rm g}$}}
\newcommand{\Pisig}{{$\Pi_{\rm sig}$}}
\newcommand{\wgstarl}{{$\mathcal{W}_{\rm g}^\star$}}
\newcommand{\wgstarlt}{{$\tilde{\mathcal{W}}_{\rm g}^\star$}}
\newcommand{\wgl}{{$\mathcal{W}_{\rm g}$}}

\newcommand{\Pismin}{{$\Pi_{\rm s,min}$}}
\begin{document} 
\title{Probing the Cores of Subdwarf B Stars: How do They Compare to Cores in Helium Core-Burning Red Giants?
}
 \titlerunning{Cores of sdB stars}
%\title{Buoyancy Glitches in pulsating stars revisited}

% \subtitle{I. Overviewing the $\kappa$-mechanism}
% The list of authors, and the short list which is used in the headers.
% If you need two or more lines of authors, add an extra line using \newauthor
\author{Margarida S. Cunha\inst{1} \and  Juliana Amaral\inst{1,2} \and Sofia Avelino\inst{2} \and Anselmo Falorca\inst{3} \and Yuri Damasceno\inst{1,2}\and Pedro Avelino\inst{1,2}}
% List of institutions
 \institute{Instituto de Astrof\'{\i}sica e Ci\^{e}ncias do Espa\c{c}o, Universidade do Porto, CAUP, Rua das Estrelas, PT4150-762 Porto, Portugal\\ \email{mcunha@astro.up.pt} \and Departamento de Física e Astronomia, Faculdade de Ciências, Universidade do Porto, Rua do Campo Alegre 687, PT4169-007 Porto, Portugal \and Leiden Observatory, Leiden University, PO Box 9513, NL-2300 RA Leiden, The Netherlands
}

% These dates will be filled out by the publisher

\date{Accepted XXX. Received YYY; in original form ZZZ}

% Enter the current year, for the copyright statements etc.
%\pubyear{2015}

% Don't change these lines
%\begin{document}
%\label{firstpage}
%\pagerange{\pageref{firstpage}--\pageref{lastpage}}
%\maketitle

% Abstract of the paper
%\begin{abstract}
\abstract
{The mixing of material from stellar convective cores into their adjacent radiative layers has been a matter of long-standing debate. Pulsating subdwarf B stars offer excellent conditions to advance our understanding of this problem. In this work we use a model-independent approach to infer information about the cores of three subdwarf B stars and compare it with similar inferences from earlier analysis of red giants in the helium core-burning phase. This is achieved by fitting an analytical description of the gravity-mode pulsation periods to pulsation data collected by the Kepler satellite. From the fits we infer the reduced asymptotic period spacings and the amplitude and position of sharp structural variations associated with chemical discontinuities in the stellar interiors. {Our results indicate the presence of sharp structural variations with similar properties in all three stars, located near the edge of the gravity-mode propagation cavity and likely associated with the C-O/He transition. We find that these structural variations differ systematically from those of helium core-burning red giant stars,  having larger amplitudes and being located at a larger buoyancy radius. This suggests that chemical mixing beyond the adiabatically stratified core into the radiatively stratified layers may be more extensive in subdwarf B stars than in helium core-burning red giants. Alternatively, the stratification of the mixing region beyond the adiabatically stratified core may differ significantly between the two types of stars. The model-independent constraints set on the structural variations inside these three stars are the first of a kind and will be key to enhance the modelling of layers adjacent to stellar convective cores and to test non-canonical stellar evolution channels leading to the formation of hot subdwarf stars. }
}

%\end{abstract}

% Select between one and six entries from the list of approved keywords.
% Don't make up new ones.
\keywords{stars: evolution -- stars: interiors -- Asteroseismology  }
\maketitle
%%%%%%%%%%%%%%%%%%%%%%%%%%%%%%%%%%%%%%%%%%%%%%%%%%

%%%%%%%%%%%%%%%%% BODY OF PAPER %%%%%%%%%%%%%%%%%%

%%%%%%%%%%%%%%%%% BODY OF PAPER %%%%%%%%%%%%%%%%%%

\section{Introduction}
\label{sec:introduction}
Subdwarf B (sdB) stars are compact, hot objects with a helium-burning core and a thin hydrogen-rich envelope that is incapable of sustaining nuclear reactions \citep[see][for a review]{heber16}. With about half a solar mass, they resemble stripped versions of low-mass helium core-burning red giant stars. Thus comparative asteroseismic studies of sdB stars and helium core-burning red giants hold the potential to significantly advance our understanding of the processes leading to the formation of sdB stars. They are often observed to be in close binaries, mostly with a white dwarf companion and thought to be a product of binary evolution involving a common-envelope phase. However, different processes, including stellar-planet interactions, stellar mergers,  and non-standard single star evolution, are also evoked as possible channels to form sdB stars \citep{clausen11,heber16,zhang17,rui24}.

A significant fraction of sdB stars exhibit either acoustic (p) or  gravity (g) modes, and some exhibit both \citep[e.g.][]{baran23,uzundag24}. Unlike the mixed modes observed in red giants, the purely acoustic or gravity nature of pulsations observed in sdB stars simplifies the development of inference methodologies. Of particular interest, the seismic data on several sdB stars have revealed signatures of sharp structural variations in the g-mode part of the pulsation spectra \citep{ostensen14,baran17,uzundag17}. These are thought to be associated with strong chemical gradients inside the star, located at the transition between the Carbon-Oxygen-enriched mixed core and the helium mantle (C-O/He transition) and between the helium mantle and the thin hydrogen-rich envelope (He/H transition).  {While both chemical transitions impact the pulsation periods, their relative seismic importance depends on the range of periods observed, as well as on the mixing processes taking place near the core \citep{Ghasemi17,guyot2025}}. Moreover, the seismic impact of the chemical transitions depends on their sharpness, in particular on how the scale of the transition compares to the local wavelength of the observed modes. This has been illustrated by \cite{charpinet14} for a sdB model, where the authors show that the seismic signature of the relatively wide He/H transition decreases significantly as the pulsation periods increase. The same phenomena was addressed in \cite{cunha15} and is discussed further in Sect.~\ref{sec:model} of this work. 

Asteroseismic studies of sdB pulsators have so far been based on forward modelling \citep[e.g.][]{Ghasemi17,uzundag21,baran23b} recurring to either stellar evolution codes or static structural models \citep{charpinet00, charpinet02,guyot2025}. However, the pure nature (gravity or acoustic) of the modes observed in these stars makes them key targets to employ model-independent asteroseismic techniques. This is particularly true for the study of sharp structural variations such as those discussed above. 

Structural variations on scales comparable or shorter than the local wavelength of the modes (hereafter, glitches) have been the focus of many seismic studies. Some of the tools for the direct inference of the properties of these glitches without recourse to stellar models date from the earlier days of helioseismology, and have seen diverse applications over the years \cite[see][for a review]{cunha20}. 
However, the variational approach adopted in the context of the sun and other p-mode pulsators to derive the analytical formulation needed to fit the seismic data and make the intended inferences is usually  not adequate for studies of g-mode pulsators. This is because the perturbations induced on the g-mode pulsation periods by glitches in the buoyancy frequency are rarely small. The same is true in the study of buoyancy glitches in mixed-mode pulsators. 

When the glitch impact on the pulsation periods is not small, perturbations to the pulsation periods may be derived instead by matching the wave solutions on each side of the glitch. This approach was used by \cite{cunha15,cunha19,cunha24} to derive analytical expressions for the period and period-spacing perturbations induced by buoyancy glitches of different shapes in g- and mixed-mode pulsators. These tools have recently been successfully applied in the analysis of glitches in a sample of low mass, helium core-burning red giant stars by \cite{vrard22}. In this work we will apply them to the study of three sdB pulsators in which evidence for glitch signatures have been detected, and compare the properties inferred for their glitches with those inferred in the work by \cite{vrard22} in the analysis of the helium core-burning red giant stars.  

The remainder of the paper is organized as follows. In Sect.~\ref{sec:data}, we present the published pulsation periods used in this work and their sources. In Sect.~\ref{sec:ps}, we introduce the analytical model for the seismic signature of structural glitches and discuss the role of each model parameter. In Sect.~\ref{sec:inferences}, we describe the fitting methods adopted and the results from the fits to the data on each star. In Sect.~\ref{sec:discussion}, we discuss our inferences in light of the combined analysis of the three stars and compare them to the results from earlier analyses of the helium core-burning red giant stars. Finally, in Sect.~\ref{sec:conclusions}, we summarize our main conclusions.

\section{Data}
\label{sec:data} 
We consider three stars for our case study, namely KIC~10553698A, studied by \cite{ostensen14} based on Kepler short-cadence data collected over 9 quarters in the period between Q8 and Q17, EPIC~211779126, studied by  \cite{baran17} based on short-cadence data collected during the K2 Campaign 5, and KIC~10001893, studied by \cite{uzundag17} based on Kepler short-cadence data collected continuously between Q6 and Q17.2. Of relevance for our study are the high radial order g modes. These are listed in Tables~\ref{tab:KIC1} and \ref{tab:KIC1_all} for KIC~10553698A, \ref{tab:EPIC} for EPIC~211779126, and \ref{tab:KIC2} for KIC~10001893, in appendix~\ref{apA}. The tables show the mode radial order, $n$, degree $l$ and period, $P$, according to the information provided in the original studies, as well as the reduced periods, defined as $\Pi=\sqrt{(l(l+1)}P$. For EPIC~211779126 and KIC~10001893, no obvious multiplets are observed in the power spectrum in the range of periods considered in this study.  However, in the case of KIC~10553698A, multiplets are observed. In Table~\ref{tab:KIC1} we list estimates of the central components of these multiplets, reproduced following the discussion in section 3.8 of \cite{ostensen14} and the authors' figure 10. The original list of mode periods in the range of relevance,  including the individual multiplet components listed in their work, is reproduced in Table~\ref{tab:KIC1_all}.  In a few cases, the three works report period spacings that were not computed from the modes listed in their tables. Estimates of the respective periods are also given in the respective tables.

\section{Periods and period spacing models in the presence of glitches}
\label{sec:ps}

In the absence of fast rotation and other agents that may break the spherical symmetry of the star, an asymptotic analysis predicts that high-radial-order gravity modes of the same degree are equally spaced in period \citep{tassoul80}. Written in terms of the reduced period, this spacing is
\begin{equation}
\Delta\Pi_{\rm as}= \frac{2\pi^2}{\mathcal{W}_{\rm g}}.
\label{psasymp}
\end{equation}
Here, \wgl\footnote{With the exception of the corner plots shown in appendix~\ref{apC}, throughout this manuscript we will use degree-independent buoyancy measurements rather than their degree-dependent counterparts 
used in earlier works by \cite{cunha15,cunha19,cunha24} for which only modes of $l=1$ were considered. The relation between the two definitions is a $\sqrt{l(l+1)}$ factor (\egc $\omega_{\rm g}\equiv\mathcal{W}_{\rm g}\sqrt{l(l+1)}$).
}    
is the size of the g-mode propagation cavity expressed in terms of the buoyancy frequency $N$ as
\begin{equation}
\mathcal{W}_{\rm g}=\int_{r_1}^{r_2}\frac{N}{r}\rmd r,
\label{eq:buoyancy_tot}
\end{equation}
where $r$ is the distance from the stellar centre, and $r_1$ and $r_2$ are the inner and outer turning
points of the cavity, respectively. { Moreover, if the star has a convective core, in the asymptotic regime the reduced periods can be written as}
\begin{equation}
    \Pi_{\rm s} \approx \Pi_{\rm s,min}+k\Delta\Pi_{\rm as},
    \label{eq:per_asymp}
\end{equation}
for a series of natural numbers $k$, where $\Pi_{\rm s,min}$ is the first of a series of equally spaced reduced periods.

We note that the reduced periods in Eq.~(\ref{eq:per_asymp}) do not depend on mode degree. In stars with a radiative core, $N^2$ is positive in the innermost layers, approaching zero only at the centre, as $N\propto r$. As a consequence, the reduced periods depend on mode degree. However, this is not expected in sdB stars, because their cores are unstable to convection, preventing g modes from propagating into the centre of the star.  In these stars the inner turning point, $r_1$, is located at the edge of the convective core, here defined as the border of the adiabatically stratified region, hence including any extra mixing regions resulting from processes that maintain an adiabatic thermal stratification \cite[see, \eg][for a discussion]{guyot2025}. Moreover, while the outer turning point, $r_2$ may generally depend on the mode degree and oscillation period, changes to the outer turning point have negligible impact on the integral defined in Eq.~(\ref{eq:buoyancy_tot}), hence on the reduced periods. That is corroborated by the data on the three stars considered in this work. In fact, the reduced periods of dipole and quadrupole modes are generally closely aligned in these stars and there is no evidence for a dependence of the asymptotic reduced period spacing on the oscillation period. 

\subsection{Glitch signature}
The equally-spaced sequence reported above is modified in the presence of structural variations occurring on scales smaller or comparable to the local wavelength (hereafter, named glitches) located within the propagation cavity. To describe the glitch position we introduce the degree-independent buoyancy radius
\begin{equation}
    \tilde{\mathcal{W}}_{\rm g}^r=\int_{r_1}^{{r}}\frac{N}{r}\rmd r,
    \label{eq:bradius}
\end{equation}
and the degree independent buoyancy depth $\mathcal{W}_{\rm g}^r=\mathcal{W}_{\rm g}-\tilde{\mathcal{W}}_{\rm g}^r$.  In all cases,  we will characterise the glitch position in terms of the buoyancy distance measured from the closest turning point.

For a single glitch, the periods are represented by \citep{cunha24}
\begin{equation}
\Pi\approx\Pi_{\rm s}-\frac{\Delta\Pi_{\rm as}}{\pi}\phi,
	\label{eq:per_1g}
\end{equation}
where $\Pi_{\rm s}$ represents the unperturbed reduced periods ({\ie} without the glitch effect) of a hypothetical model in which the integral of $N/r$ within the g-mode cavity is the same as in the glitch model, and $\phi$ represents the phase perturbation caused by the glitch.  Moreover, the glitch-induced phase depends on the functional form of the glitch. Here, two options shall be considered, namely a step-like function and a Gaussian-like function (see figure 1 of \cite{cunha19} for illustration). 
For the case of the a step-like glitch, the phase is given by
\begin{equation}
    \phi={\rm arccot}\left[-\frac{2}{A_{\rm st}\sin\left(2\theta\right)}-\cot\theta\right],
	\label{eq:phase_step}
\end{equation}
while in the case of a Gaussian-like glitch, the  phase is given by
\begin{equation}
    \phi={\rm arccot}\left[\frac{1}{A_{\rm G}{\it f}_{\omega}^{\Delta_{\rm g}}\sin^2\theta}-\cot\theta\right],
        \label{eq:phi_gau}
\end{equation}
where, $f_{\omega}^{\Delta_{\rm g}}=\omega^{-1}{\exp}(-2\Delta_g^2\omega^{-2})$ and $\omega$ is the angular frequency.
Here, \as\ and \ag\ are always positive and provide a measure of the amplitude of the glitch located at $r=r^\star$, in the case of the step-like and Gaussian-like glitch, respectively. 
Moreover, $\Delta_{\rm g}$ is a measure of the width of the Gaussian-like glitch and 
\begin{equation}
    \theta=\frac{\tilde{\mathcal{W}}_{\rm g}^\star}{2\pi}\Pi+\delta+\frac{\pi}{4} \mbox{ \;\;or\;\; } \theta=\frac{\mathcal{W}_{\rm g}^\star}{2\pi}\Pi+\delta+\frac{\pi}{4} ,
	\label{eq:teta}
\end{equation}
depending on whether the glitch is located in the inner half (\wgstarlt/\wgl<0.5) or outer half (\wgstarlt/\wgl>0.5) of the propagation cavity, respectively. We note that in the above, a $\star$ means that the quantities are taken at $r=r^\star$.
 Finally, $\delta$ is an extra phase that absorbs the impact of the approximations performed in the analysis near the turning points of the modes and is not related to the glitch shape. This extra parameter also accommodates the impact in Eq.~(\ref{eq:phase_step}) of changing the sign of the structural variation (\eg a local drop versus a local increase in buoyancy frequency) in the case of a step-like glitch. This sign change will introduce a $\pi/2$ variation in the parameter $\delta$, without changing the expression itself.  By symmetry, this also means that the expression for the phase $\phi$ is independent of the side of the cavity in which the glitch is located,  insofar as the correct definition for the buoyancy position of the glitch (buoyancy radius or depth, as discussed above) is taken \cite[see][for details]{cunha24}. 
 
 Finally, we note that Eq.~(\ref{eq:phase_step})
 is periodic, with a period given by
 \begin{equation}
   \Pi_{\rm sig} =  \frac{2\pi^2}{\tilde{\mathcal{W}}_{\rm g}^\star} \mbox{ \;\;and\;\; }  \Pi_{\rm sig} =  \frac{2\pi^2}{\mathcal{W}_{\rm g}^\star},
 \end{equation}
 for (\wgstarlt/\wgl<0.5) and (\wgstarlt/\wgl>0.5), respectively.
That implies that the signature of a step-like glitch on the reduced periods will be strictly repeated with this periodicity. In the case of a Gaussian-like glitch, inspection of Eq.~(\ref{eq:phi_gau}) shows that the glitch signature also changes periodically with the same periodicity, but its amplitude is simultaneously modified due to the frequency-dependent function $f_{\omega}^{\Delta_{\rm g}}$.

The analysis followed by these authors can be expanded to consider additional buoyancy glitches, so long as these are sufficiently apart to guarantee that the asymptotic solutions are valid in between them (see Appendix~\ref{apB}). In the case of two glitches, the reduced periods are given by,
\begin{equation}
\Pi\approx\Pi_{\rm s}-\frac{\Delta\Pi_{\rm as}}{\pi}\left(\phi+\varphi\right),
	\label{eq:per_2g}
\end{equation}
where $\varphi$ represents the phase perturbation caused by the additional glitch.  As in the case of the first glitch, the functional form of $\varphi$ is given either by Eq.~(\ref{eq:phase_step}) or Eq.~(\ref{eq:phi_gau}), depending on whether it is modelled by a step or a Gaussian function, respectively. Moreover, if the two glitches are located in opposite sides of the g-mode cavity, $\theta$ is given by the left and right expression of Eq.~(\ref{eq:teta}), for the inner and outer glitch, respectively. Finally, the expression for the period spacings,  defined by the difference in the reduced period of modes of consecutive radial order $\Delta\Pi=\Pi_{n+1}-\Pi_n$, is given by
\begin{equation}
    \frac{\Delta\Pi}{\Delta\Pi_{\rm as}}\approx\left[1+\frac{\Delta\Pi_{\rm as}}{\pi}\left(\frac{\rmd\phi}{\rmd\Pi}+\frac{\rmd\varphi}{\rmd\Pi}\right)\right]^{-1},
	\label{eq:quadratic}
\end{equation}
where the derivatives of the phases $\phi$ and $\varphi$ can be computed numerically  or analytically \cite[\cf][]{cunha19}. 

\subsection{Model parameters}
\label{sec:model}
When fitting the analytical expression (\ref{eq:quadratic}) to data, the number and type of parameters depend on the number of glitches and glitch shapes. The common parameter to all cases is the asymptotic reduced  period spacing, \dpias. Then, for each glitch we have the buoyancy radius \wgstarlt\, (or the buoyancy depth, \wgstarl) at the glitch position, and the fudge parameter \phase. Depending on the shape of the glitch, we have also either a dimensionless amplitude, \as\ (step-like glitch) or an amplitude and a width, \ag\ and \width, respectively (Gaussian-like glitch). When fitting the analytical expression for the periods (\ref{eq:per_2g}), there is an additional parameter that represents the smallest reduced period in the fitted range, \Pismin.  

  \begin{figure*}[h!]
    \includegraphics[width=2\columnwidth]{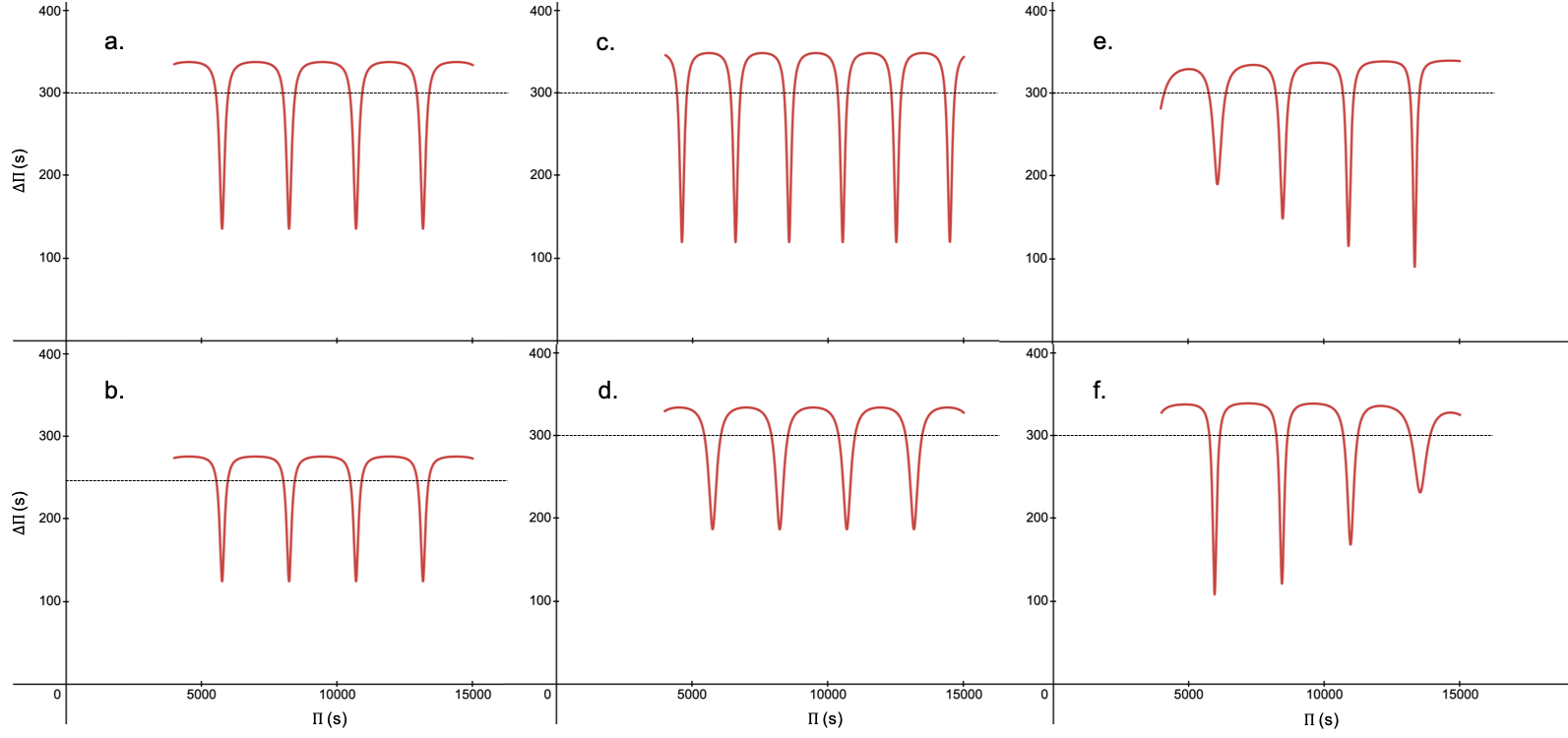}
    \caption{Continuous reduced period spacing signal predicted by the analytical model (Eq.~(\ref{eq:quadratic})), considering a single step-like glitch (Eq.~(\ref{eq:phase_step})) and varying the model parameters one at a time (red line). The horizontal dotted line marks the asymptotic reduced period spacing value \dpias. {Panel a.}: reference case with \dpias =300~s, \as =10, \wgstarlt=0.008 rad/s (or \Pisig =2467~s) and \phase =1.3. {Panel b.}: same as reference but with \dpias=250~s. {Panel c.}: same as reference but with \wgstarlt=0.010~rad/s (or \Pisig=1974~s). {Panel d.}: same as reference but with \as=5. {Panel e.}: same as reference case but for a Gaussian glitch with \ag=0.002~rad/s and \width=0~rad/s. {Panel f.}: same as reference case  but for a Gaussian glitch with \ag=0.006~rad/s and \width=0.00049~rad/s.}
    \label{fig:desmos}
\end{figure*}

Figure~\ref{fig:desmos} illustrates the impact on the period spacings from changing the physical parameters of a single-glitch model, as well as from changing the glitch shape. The reference model, shown in panel a., considers \dpias~=~300~s and a step-like glitch with \as~=~10, \wgstarlt~=~0.008 rad/s (or \Pisig~ =~2467~s) and \phase~=~1.3. These parameters imply that the relative position of the glitch in the cavity for the reference model is \wgstarlt$/$\wgl$\approx 0.12$. 
\begin{itemize}[label=$-$]
    \item The asymptotic reduced period spacing, \dpias, controls the value around which the the period spacing deviations take place. In addition, it acts as a scaling factor for the glitch-induced perturbations, as expected from Eq.~(\ref{eq:per_2g}). This impact is illustrated in panel b.\ where the value has been decreased to \dpias~=~250~s.
    \item The glitch position, \wgstarlt\ (or \wgstarl, for a glitch in the outer cavity), controls the separation between the extrema of the period-spacing function, with larger values implying smaller separations. This is shown in panel c., where the glitch was moved slightly closer to the centre of the cavity, with \wgstarlt=0.01~rad/s (\wgstarlt$/$\wgl$\approx 0.15$; \Pisig=1974~s). 
    \item The glitch amplitude, \as\ for a step-like glitch, controls the extent of the period-spacing deviation from \dpias, with a smaller amplitude (\as=5; panel d.) showing shallower humps and dips. When the amplitude is sufficiently small, the variation of the reduced period spacing becomes sinusoidal and symmetrical about the value of \dpias. 
    \item Changes in the phase, \phase, induce only a horizontal translation of the reduced period spacings (not shown in Fig.~\ref{fig:desmos}).
    \item Changing the shape of the glitch impacts the dependence of the reduced period-spacing perturbations on the reduced periods. While for a step-like glitch the extrema of the perturbations are independent of the reduced period (panels a. to d.), in a Gaussian-like glitch the period dependence is clear (panels e. and f.). This dependence is controlled by the function  $f_{\omega}^{\Delta_{\rm g}}$ in Eq.~(\ref{eq:phi_gau}). When the width of the glitch, \width, tends to zero, the period dependence is dominated by the factor $1/\omega$ and the extrema of the perturbations increase with the reduced period (panel e.), otherwise, the dependence is dominated by the exponential factor and the extrema of the perturbations decrease as the reduced period increases (panel f.). The latter behaviour results from the fact that as the reduced period increases, the local wavelength of the wave decreases, eventually becoming smaller than the width of the structural variation. At that point, the structural variation ceases to be a glitch and the glitch-induced perturbation disappears.
\end{itemize}

\section{Inference of glitch properties}
\label{sec:inferences}
The reduced period spacings are shown for the three stars in the top panels of Figs~\ref{fig:KIC1}, \ref{fig:EPIC}, and \ref{fig:KIC2}, respectively. In all three stars, reduced period-spacing dips are observed, consistent with the signature of structural glitches.  
We note that the reduced period spacings for the $l=1$ and $l=2$ modes are generally aligned, as expected asymptotically for a star with a spherically symmetric equilibrium and a convective core \citep[\eg][]{cunhaetal07}. Exceptions to this alignment may indicate problems with the identification of the modes, as noted in the original works. These cases shall be discussed further below.

To fit the period-spacing model to the data and infer the glitch properties, a Dynamic Nested Sampling algorithm \citep{Higson_2018} was implemented using the dynesty pyhton package.

The likelihood was assumed to be Gaussian and given by the following expression
\begin{equation}
    \mathcal{L}=\left(\frac{1}{\sqrt{2\pi \sigma^2}}\right)^N \exp\left(-\frac{1}{2}\chi^2\right),
	\label{eq:likelihood}
\end{equation}
where $\chi^2$ is defined as:
\begin{equation}
    \chi^2=\sum_i^N\left(\frac{ \mathcal{Q}_{{\rm data,} i}-\mathcal{Q}_{{\rm model,}i}}{\sigma}\right)^2,
	\label{eq:chisquare}
\end{equation}
$N$ is the number of data points to be fitted and $\mathcal{Q}_{{\rm data}}$ and $\mathcal{Q}_{{\rm model}}$ are the data and respective model quantity being compared  (in the present case, the period spacings $\Delta\Pi$). In addition to the parameters implicit in the analytical model for the periods and period spacings, equation (\ref{eq:chisquare}) contains the error, $\sigma$, that is also taken to be a free parameter. We note that no observational errors have been provided in the original works from which the data used here were extracted, so using the observational errors in the definition of the likelihood is not an option. In any case, we anticipate that the typical observational errors - connected to the pulsation spectra resolution, mode amplitude variability, etc.- are negligible compared with the errors associated with the model predictions. In fact, while the analytical expressions are expected to capture the essence of the glitch impact on the pulsation periods, it is likely that in most cases they are insufficiently accurate to model the data within typical observational errors. This is due both to the approximations inherent to the analysis performed to derive the analytical expressions and to the fact that glitches in real stars are not expected to be perfectly described by step, Gaussian or Dirac $\delta$ functions. Consequently, the introduction of $\sigma$ as a parameter enables us not only to proceed without having access to the observational errors but also to quantify the ability of our models to reproduce the data. 

We ran the nested sampler algorithm with 250 live points and a burn-in set to one-tenth of the total chain length, using multiple bounding ellipsoids and setting the sampling method to be random walks. Uniform priors were used for all parameters within the parameter space explored. From these, we extracted the parameters corresponding to the maximum of the likelihood, as well as the median parameters of the posterior distribution and associated 68~per cent confidence intervals. 

In addition, for KIC~10553698A we fitted the reduced periods using a step-like glitch (\iec Eqs~(\ref{eq:per_1g})-(\ref{eq:phase_step})) and different options for the data selection, as detailed in Sect.~\ref{sec:kic1}. This was motivated by the presence of multiplets in the g-mode range in this star, which makes the identification of the central peaks needed to construct the period spacings particularly difficult \citep{ostensen14}. The fits to the reduced periods followed the procedure described in \cite{cunha24}, where the data were fitted through a Markov Chain Monte Carlo (MCMC) method using the emcee python package and the likelihood was also assumed to be Gaussian (Eqs~(\ref{eq:likelihood})-(\ref{eq:chisquare}) with $\mathcal{Q}=\Pi$) and the priors were taken to be uniform within the parameter space explored. 

For each star, several scenarios were tested. The motivation for the choice of glitch models and the results from the fits for each star are discussed in the following subsections.

\subsection{KIC~10553698A}
\label{sec:kic1}
\begin{figure}
 	\includegraphics[width=\columnwidth]{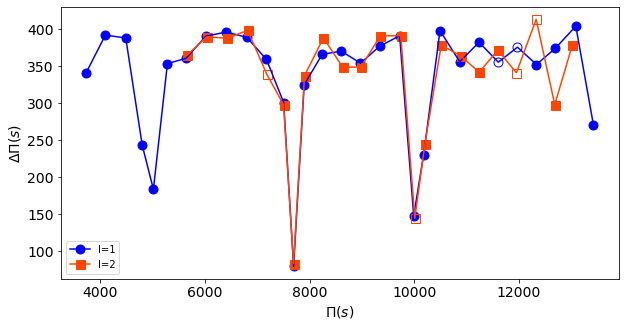}
	\includegraphics[width=\columnwidth]{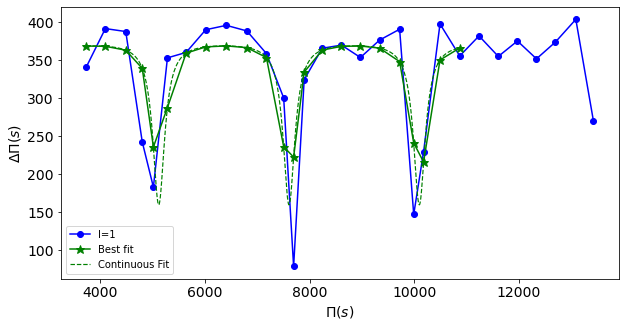}
		\includegraphics[width=\columnwidth]{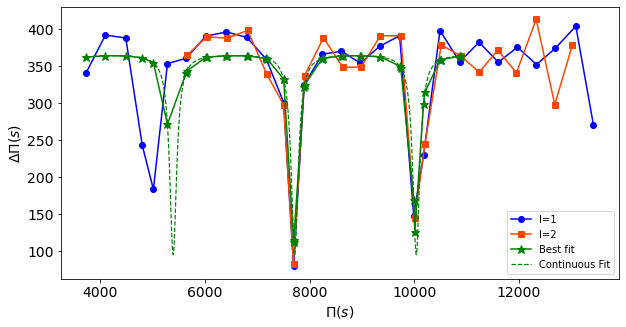}
    \caption{KIC~10553698A reduced period spacings as a function of the reduced period. Each reduced period spacing is plotted at the mid point between the reduced periods used in  its computation. The parameters for the best fit solutions are given in Appendix~\ref{apC}. {Top panel:} observed reduced period spacings for $l=1$ (blue) and $l=2$ (red) in the full range. The discrete observed values are joined by straight lines to guide the eye. Open symbols indicate reduced period spacings that cannot be computed from the frequencies listed in the original table \cite[see section 3.8 of][for details]{ostensen14}. {Middle panel:} the best fit solution to the $l=1$ reduced period spacings at reduced periods smaller than 11100~s, for the case of a one step-like glitch (green) (case A).  The continuous analytical solution is represented by the dashed line, while the filled stars indicate the values of the solution at the reduced periods corresponding to the locations of the observations. {Bottom panel:} similar to middle panel, but including in the fit both the $l=1$ (blue) and the $l=2$ (red) modes at reduced periods smaller than 11100~s (case B).}
    \label{fig:KIC1}
\end{figure}

 Two modulations appear to be present in the reduced period spacings of KIC~10553698A, with different amplitudes and scales,  (Fig.~\ref{fig:KIC1}, top panel). The first, of a larger scale, is characterised by three significant dips, while the second, of shorter scale, has a comparatively small amplitude. Although both modulations seem to be present and in phase in the modes of degree $l=1$ and $l=2$ up to a reduced period of $\Pi\sim 11100$~s, the short-scale modulation on the the two mode degrees becomes out of phase for reduced periods larger than that value. As discussed by \cite{ostensen14}, amplitude variability and rotational splitting make the identification of the modes difficult in this star. This is particularly so at high radial orders (large periods) where the peaks in the multiplets become increasingly apart in period. For this reason, the authors state that the offsets seen in the reduced period spacings computed from modes of different degree are likely not of physical origin. With this in mind, for KIC~10553698A we disregard the reduced period spacings for periods above $11100$~s when fitting our analytical model. Moreover, given that the amplitude of the short-scale modulation is similar in magnitude to the differences in the reduced period spacings for $l=1$ and $l=2$ modes, we disregard also that modulation when choosing the model to fit. Therefore, we  consider a one glitch model aimed at fitting the dominant larger-scale modulation observed in the reduced period spacings of KIC~10553698A. The absence of a significant change in the amplitude of the glitch signature, which shows variations that can be well explained by the limited sampling by the data,  points to a sharp transition, such as the one that may be expected in the C-O/He transition (\cf Sect.~\ref{sec:introduction}). Thus, we shall assume a step-like glitch, which provides the simplest possible representation of the data, with a minimum number of model parameters. 
 
  \begin{table*}[h!]
	\caption{Seismic and glitch properties of KIC~10553698A derived from fits of a step-like glitch model to the data.}
	\label{tab:KIC1_parameters}
 \centering
\begin{minipage}{0.95\textwidth}
	\resizebox{\linewidth}{!}{%
  \begin{tabular}{l c c c c c|c l c c}
\hline																			
\hline																			
\rule{0pt}{3ex}																			
	&	$\Pi_{\rm s,min}$ (s)	&	$\Delta\Pi_{\rm as}$ (s) 	&	 $A_{\rm st}$  	&	$\tilde{\mathcal{W}}_{\rm g}^\star$ ( rad/s) 	&	 $\delta$ 	&	$\Pi_{\rm sig}$ (s)	&	 model and data	&	range	&	case	\\[1ex]
\hline																			
\rule{0pt}{4ex} 																			
Period Spacings' fit	&		&	$306^{+28}_{-18}$	&	$40^{+51}_{-34}$	&	$0.0086^{+0.0004}_{-0.0012}$	&	$1.2^{+1.5}_{-0.4}$	&	$2295^{+320}_{-107}$	&	Step ; $l=1$	&	<11100 s	&	A	\\[4ex]
	&		&	$317^{+20}_{-14}$	&	$17^{+37}_{-10}$	&	$0.0085^{+0.0003}_{-0.0006}$	&	$1.3^{+0.9}_{-0.3}$	&	$2322^{+164}_{-82}$	&	Step ; $l=1,2$ 	&	<11100 s	&	B	\\[2ex]
\hline																			
\rule{0pt}{4ex} 																			
Periods' fit	&	$3616^{+14}_{-14}$	&	$325.42^{+0.96}_{-0.95}$	&	$55^{+29}_{-24}$	&	$0.00785^{+0.00012}_{-0.00012}$	&	$0.69^{+0.18}_{-0.17}$	&	$2515^{+38}_{-38}$	&	Step ; $l=1,2$ 	&	<11100 s	&	C	\\[4ex]
\rule{0pt}{4ex} 	&	$3622^{+24}_{-24}$	&	$325.5^{+1.7}_{-1.7}$	&	$62^{+29}_{-26}$	&	$0.00778^{+0.00019}_{-0.00022}$	&	$0.87^{+0.23}_{-0.30}$	&	$2537^{+62}_{-72}$	&	Step ; $l=1,2$ ; random mean	&	<11100 s	&	D	\\[4ex]
\rule{0pt}{4ex} 	&	$3589^{+27}_{-27}$	&	$328.5^{+1.7}_{-1.6}$	&	$37^{+39}_{-22}$	&	$0.00707^{+0.00018}_{-0.00020}$	&	$1.62^{+0.30}_{-0.26}$	&	$2792^{+79}_{-71}$	&	Step ; $l=1,2$ ; random	&	all	&	E	\\[2ex]
 \hline																													
 \end{tabular}
  }
\end{minipage}
\tablefoot{Parameters derived from the fits of a model of a step-like glitch to the reduced period spacings (cases A and B) and to the reduced periods (cases C, D, and E) of KIC~10553698A (see text for details). The results on the left side of the table show the medians and the 68~per~cent confidence intervals of the inferred parameters, from left to right, the first of the reduced periods in the range considered, the asymptotic reduced period spacing, the glitch amplitude and position, and the phase. The right side of the table indicates the periodicity of the glitch signature and the choice of model and data.}
\end{table*}

%-------------------------------------------------------------

 Two fits were performed to the reduced period spacings. The first considers only the modes of $l=1$, which cover all observed dips (case A). Fitting to the $l=1$ data thus ensures that all dips are given the same weight. The best-fit solution for this case (\ie the solution with maximum likelihood) is shown in the middle panel of Fig.~\ref{fig:KIC1} (green stars), along with the observed $l=1$ reduced period spacings (blue circles). Here, and in all following figures, the solutions are shown only in the range of reduced period spacings where the fits are performed. The solution shows that the model captures well the three-dip structure seen in the observed period spacing. This is corroborated by the posterior distribution for the glitch position shown in Fig~\ref{fig:corner_KIC1_1} (in appendix~\ref{apC}). The latter shows that the model parameter that measures the buoyancy radius at the glitch position, that affects more directly the distance between period-spacing dips, is well constrained. The fact that the extremes of the dips are not well reproduced by the best-fit solution (\cf Fig.~\ref{fig:KIC1}, middle panel) is not surprising, as the dips are narrow and not well sampled by the discrete data. As a consequence, the fraction of data points contributing to constraining the dips is relatively small.  In fact, considering that the phase $\delta$ simply induces a horizontal translation of the analytical solution, it is possible to see from the continuous solution for the best-fit model (shown by the green dashed line) that the extremes of the model dips will change significantly if a slightly different value of the phase is taken for a glitch of equal position and amplitude. 
 
 Unlike the glitch position, the glitch amplitude is not well constrained, as seen by inspection of the respective posterior distribution (Fig~\ref{fig:corner_KIC1_1}). The reason is again related to the limited sampling of the dips in the reduced period spacings. In the analytical model, the glitch amplitude controls the minima of the period-spacing dips, which decrease further as the amplitude increases. However, no matter how extreme the dips predicted by the continuous model are, a slight change in the phase will result in the dips being sampled away from their minima, mimicking the signal of a smaller amplitude glitch. As a result, in a poorly sampled case the glitch amplitude is never well constrained at the higher values.  Moreover, at the lower end of the amplitude posterior distribution, the large errors implied by the simplicity of the model (which assumes a single step-like function) also limit the constraining power of the data regarding the glitch amplitude due to the appearance of solutions corresponding to a flat reduced period spacing (\iec no glitch) with values smaller that the preferred solution (\iec outside the 68~per cent interval for $\Delta\Pi_{\rm as}$). These results are in line with earlier discussions in the literature in the context of both buoyancy and acoustic glitches, where it is systematically found that the glitch parameter whose inference is most reliable is the glitch position \citep{mazumdar14,cunha24}. 
 The median and associated 68~per cent uncertainty intervals for the physical parameters inferred from the fit are summarized in the first row of Table~\ref{tab:KIC1_parameters} (case A). 

 The second fit to the reduced period spacings of KIC~10553698A adopted the same model but considered both modes of degree $l=1$ and $l=2$ up to $\Pi=11100$~s (case B). The best-fit solution and the posterior distributions for the model parameters are shown in Fig.~\ref{fig:KIC1} (bottom panel) and Fig~\ref{fig:corner_KIC1_2}, respectively. Here, the increase in the number of data points leads to a better representation of the extremes of the second and third period-spacing dips, at the expense of having the first dip less well reproduced by the model. {We note that the Gaussian-glitch model also predicts equally-spaced dips in the reduced period spacing (\cf Fig.~\ref{fig:desmos}). Thus, adopting instead a Gaussian-glitch model would not solve the small mismatch between the equally spaced dips predicted by the model and the slightly uneven observed spacing. The mismatch may have an observational origin, if one of the modes near the first dip was misidentified due to the complexity of the oscillation spectra discussed above. Alternatively, it could be an indication that the glitch has a more complex struture or that the model becomes less adequate at lower periods.} 
 
 The impact of the additional (and consistent) $l=2$ data is clearly reflected in the posterior distribution for the glitch amplitude. In fact, the added weight of the data around the second and third dips removes some of the degeneracy of the model solutions corresponding to large glitch amplitudes.  Moreover, the fact that the model representation of the data is improved when adding the $l=2$ modes (as seen from the error parameter distribution), makes solutions of very low amplitude significantly less likely than when fitting only the $l=1$ modes.  These results show that if enough data are available, the glitch amplitude can be reasonably constrained, despite the large uncertainties. Comparison of the median values for the physical parameters in the two cases (Table~\ref{tab:KIC1_parameters}, cases A and B) shows that the inferred parameters are consistent within the uncertainties, with the second fit setting more stringent constraints, as expected.

 As mentioned earlier, KIC~10553698A shows a series of multiplets, or partial multiplets, in the g-mode period range. These add to the complexity of its oscillation spectra and may compromise the identification of modes of similar degree and azimuthal order, needed to determine the reduced period spacings.  For this reason, it is of interest to understand how the results from fitting directly the oscillation periods, without assuming prior knowledge of the azimuthal orders, compare with those from fitting the period spacings assuming prior knowledge of the central peaks in each multiplet. 
 
\begin{figure}
 	\includegraphics[width=1.2\columnwidth]{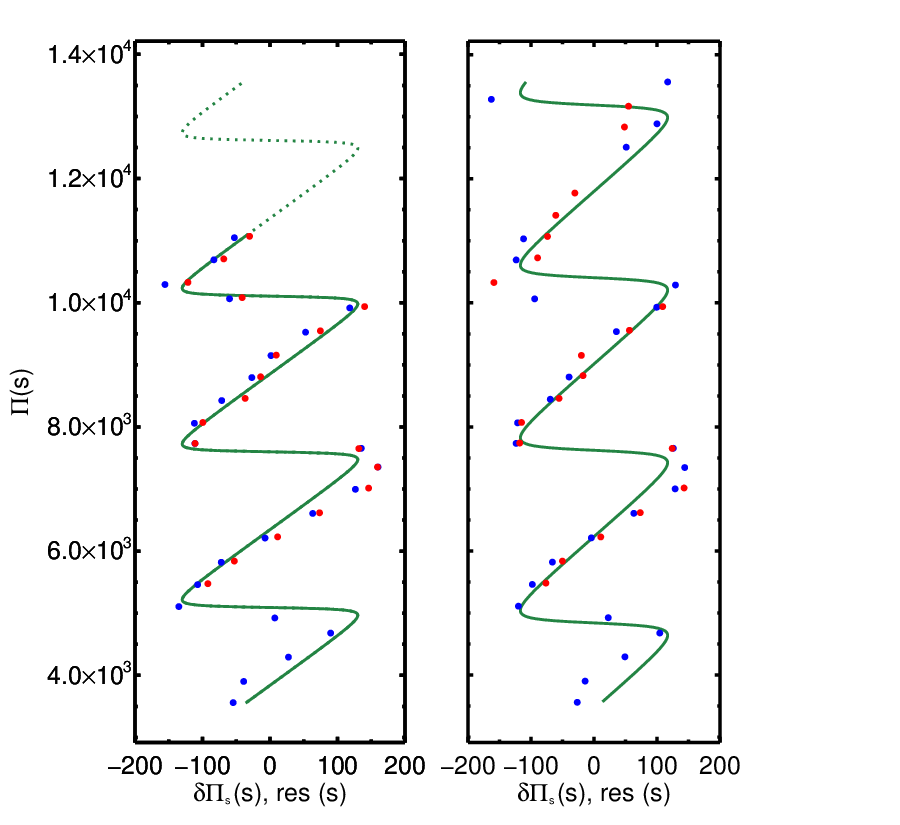}
    \caption{KIC~10553698A echelle diagrams displaying the data for $l=1$ (blue) and $l=2$ (red) modes, along with the best fit model for the case of a one step-like glitch (green). The continuous analytical solution is represented by the full line in the reduced periods' range in which the fit was performed and by a dotted line in the range not considered in the fit. {Left panel}: fit to $l=1$ and $l=2$ modes' reduced periods smaller than $11100$~s (case C). The data used in this fit were also used to construct the reduced period spacings for case B in Table~\ref{tab:KIC1_parameters}, whose fit is shown in the bottom panel of Fig.~\ref{fig:KIC1}. {Right panel:} fit to $l=1$ and $l=2$ modes' reduced periods in the full range (up to $n=34$) extracted from Table~\ref{tab:KIC1_all} (case E). When several modes are listed for a given radial order and degree (or for the same trapped mode), one single mode was picked randomly (see text for details).  }
    \label{fig:KIC1_echelle}
\end{figure}

 The results obtained from fitting the periods and the period spacings constructed from the same periods should be consistent. To verify this is the case, we start by fitting the reduced periods smaller than 11100~s listed in Table~\ref{tab:KIC1} for the $l=1$ and $l=2$ modes, used to construct the reduced period spacings fitted in the previous exercise. The reduced periods were fitted with the same step-like glitch model as before, now expressed through Eqs~(\ref{eq:per_1g})-(\ref{eq:phase_step}), and the median values for the inferred parameters are listed in Table~\ref{tab:KIC1_parameters} (case C). The glitch properties inferred from this fit agree within the errors with the results of the fit to the reduced period spacings constructed from the same data (case B). The best-fit model from the fit to the reduced periods is shown in the Echelle diagram displayed in the left panel of Fig.~\ref{fig:KIC1_echelle}, constructed in the manner described in \cite{cunha24}. 
 
 Next, we relax the assumptions made regarding the identification of the central peaks in each multiplet, used in the construction of the period spacing series \citep{ostensen14}. To that end, we consider the full list of oscillation periods published by the authors in the same range of radial orders as before (Table~\ref{tab:KIC1_all}). For each radial order and degree, we randomly choose one of the values whenever more than one period is listed. Restricting the data to reduced periods smaller than 111000~s, we then perform a fit to the reduced periods using Eqs~(\ref{eq:per_1g})-(\ref{eq:phase_step}). The exercise was repeated three times, so that the list of reduced periods fitted each time is slightly different, due to the random choice of the peak in each multiplet. For all inferred parameters, the variance in the median was found to be much smaller (always less than 3~per cent) than the errors in the respective parameter, which implies that the choice of peak in each multiplet does not affect the inferences in a significant way. The means of the parameters (\iec of the medians and respective errors) inferred from the fits to the three different data sets selected in the way described above are also listed in Table~\ref{tab:KIC1_parameters} (case D).
 
 Finally, we relax also the restriction on the range of reduced periods to be fitted, considering the full list of periods in Table~\ref{tab:KIC1_all} and randomly choosing a reduced period from each multiplet, as before. The best-fit model for this case is illustrated in the Echelle diagram displayed on the right panel of Fig.~\ref{fig:KIC1_echelle} and the inferred parameters are also listed in Table~\ref{tab:KIC1_parameters} (case E). Here we see that the reduced asymptotic period spacing and glitch amplitude agree within the errors with those inferred in cases A to D, but the glitch position is off by about 2-4 times the respective uncertainty, depending on the case. This is a consequence of the additional data fitted (\iec the reduced periods larger than 11100~s) and not of the random selection of the peaks in the multiplets or of the option to fit the reduced periods rather than the reduced period spacings.

\subsection{EPIC~211779126}
\label{baran}
\begin{figure}
	\includegraphics[width=\columnwidth]{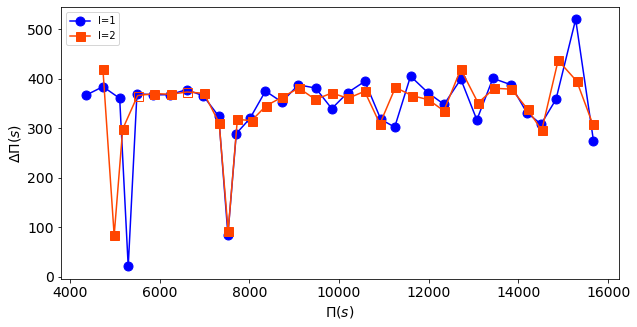}
 \includegraphics[width=\columnwidth]{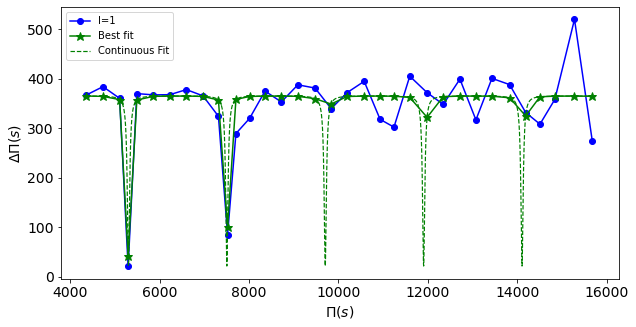}
	\includegraphics[width=\columnwidth]{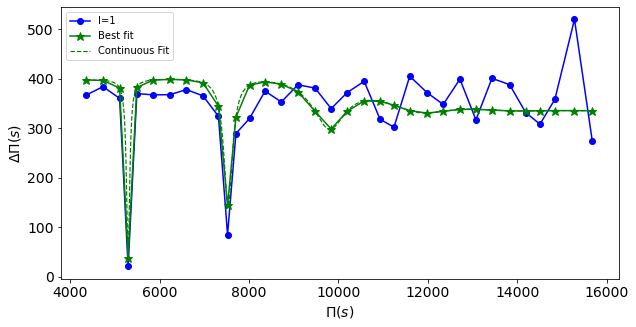}
    \caption{EPIC~211779126 reduced period spacings as a function of the reduced period. Each reduced period spacing is plotted at the mid point between the reduced periods used in  its computation. The parameters for the best fit solutions are given in Appendix~\ref{apC}.
    {Top panel:} observed reduced period spacings for $l=1$ (blue) and $l=2$ (red) modes.  The discrete values are joined by straight lines to guide the eye. Symbols meaning as in the top panel of Fig.~\ref{fig:KIC1}. {Middle panel:} the best fit solution (green) to the $l=1$ data (blue) for the case of a one step-like glitch (case F). The continuous analytical solution is represented by the dashed line, while the filled stars indicate the values of the solution at the reduced periods corresponding to the locations of the observations. {Bottom panel:} same as middle panel but for the case of a one Gaussian-like glitch (green) (case H). 
}
    \label{fig:EPIC}
\end{figure}

The reduced period spacings of EPIC~211779126 (top panel of Fig.~\ref{fig:EPIC}) show two dips between 5000~s and 8000~s, followed by smaller amplitude variations. The modes in this star also exhibit amplitude variability, a fact emphasized by \cite{baran17} with a word of caution when the authors say that the frequencies they list should be considered approximate. In fact, the reduced period spacings computed from the $l=1$ and $l=2$ data are partially in phase, particularly in the second reduced period-spacing dip and near the high end of the reduced periods, but a mismatch is seen in the first dip and at intermediate values of the reduced periods. {In particular, the separation between the dips in the reduced period spacings  is different in the $l=1$ and $l=2$ mode series. This is not expected from asymptotic analysis, and is not seen in pulsation calculations of sdB models, as far as we are aware \citep{charpinet02,Ostrowski2021}. While we do not exclude the possibility that  the differences in the two series may be real, our glitch model uses the asymptotic solutions on either side of the glitch and, thus, predicts similar results for the modes of different degree, for a given structural glitch}. For that reason, we fit the modes of different degree separately, adopting a one-glitch model, and then access the impact from the differences observed in the $l=1$ and $l=2$ data on the glitch inferences. Given our choice of a single-glitch model, we expect to reproduce only the dominant dips in the reduced period spacings. 

Concerning the representation of the glitch, we consider two cases, namely a step 
and a Gaussian function. The motivation for comparing these two cases comes from the fact that no significant dips are observed beyond $\sim~7500$~s. There are several possible explanations for this fact. On the one hand, the absence of the dips can be related to the observations themselves. With dips as extreme as the ones observed in this star, where the difference between the periods of the trapped mode and the mode that follows are as small as 20~s, and the amplitude variability already discussed, one needs to consider the possibility that some modes in the range of periods considered here were not detectable in the power spectrum. Missing modes between those used to construct the period spacings shown in the top panel of Fig.~\ref{fig:EPIC} would lead to the absence of what would otherwise be additional dips. On the other hand, the absence of additional dips can be real. 
If the structural variation is not sufficiently sharp, at larger periods its width may be significantly larger than the local characteristic scale of the wave, in which case the structural variation will no longer act as a glitch. In this case, the signature of the structural variation on the period spacings will decrease with period, tending to zero as the structural variation becomes a smooth background variation from the perspective of the wave, as discussed in Sect.~\ref{sec:model} \cite[see also discussions in][]{charpinet14,cunha20}. That is precisely what is predicted when the glitch is modelled by a Gaussian function. Finally, the glitch may be sharp, but the conditions necessary for the occurrence of visible trapping are not always satisfied. This case, as well as the cases in which it may be difficult to detect all modes in a given region of the power spectrum, is best modelled by a step function.

The best fits to the $l=1$ modes are shown in Fig.~\ref{fig:EPIC} for a model with a step-like glitch (middle panel, case F) and a model with a Gaussian glitch (bottom panel, case H). Both models capture the two-dip structure, but adjust differently to a comparably flatter reduced period spacing at higher reduced periods, as expected. In the case of the step-like glitch, the reduced period-spacing dips of the best-fit continuous solution are not always revealed by the observed modes, whereas in the Gaussian-like model, the dips disappear because the amplitude of the glitch signature decreases with reduced period. 

The posterior distributions for the model parameters and respective medians and 68~per cent uncertainties are shown in Figs~\ref{fig:corner_EPIC_1} (step-like glitch model) and \ref{fig:corner_EPIC_2} (Gaussian-like glitch model) and Table~\ref{tab:EPIC_parameters} (cases F and H). As before, in both cases we see that the glitch position is well constrained, while the glitch amplitude is not. Moreover, comparison of the median values of the glitch position inferred in the two cases shows agreement within the uncertainties, as expected from the fact that both glitch models are capable of capturing well the two dips observed in the reduced period spacings.

Fits to the $l=2$ modes were performed with the same models (cases G and I). The best-fit models adjust to the data similarly well as they did when fitting to the $l=1$ modes, so we refrain from showing the plots here. However, in Table~\ref{tab:EPIC_chi2}, we provide a measure of the quality of the fitting for the four cases (two for $l=1$ and two for $l=2$) for comparison. As the error, $\sigma$, was left as a free parameter, we cannot use $\chi^2$ defined by equation~(\ref{eq:chisquare}) directly to compare the quality between the fits performed to different sets of data or with different glitch models. Instead, we first renormalize the $\chi^2$ obtained for each fit by multiplying it by the value of $\sigma^2$ obtained for the best-fit model and then dividing the result by a reference value, namely $\sigma^2_{40}={(40\,{\rm s})}^2$. We then divide the result by $N-m$, where $N$ is the number of fitted data points and $m$ the number of parameters in the model, to compute the corresponding reduced value, $\chi^2_{R40}$, that we show in Table~\ref{tab:EPIC_chi2}. 

The medians and 68~per~cent uncertainties for the parameters inferred from the fits to the $l=2$ data are also shown in Table~\ref{tab:EPIC_parameters} (cases G and I). As expected, the position of the glitch inferred when fitting to the $l=2$ data differs from that inferred when fitting to the $l=1$ data (in that the value of $\tilde{\mathcal{W}}_{\rm g}^\star$ is smaller for $l=2$)  because the dips in the reduced period spacings are further apart in the $l=2$ case. 

%-------------------------------------------------------------
 \begin{table*}[h!]
	\caption{Seismic and glitch properties of EPIC~211779126 derived from fits of step-like and Gaussian glitch models to the data.
}
 \centering
\label{tab:EPIC_parameters}
\begin{minipage}{0.8\textwidth}
	\resizebox{\linewidth}{!}{%
  \begin{tabular}{l c c c c |c c l c}
\hline																		
\hline																		
	\rule{0pt}{3ex}$\Delta\Pi_{\rm as}$ (s) 	&	 $A_{\rm st}$  or  $A_{\rm G}$  (rad/s)	&	$\tilde{\mathcal{W}}_{\rm g}^\star$ ( rad/s) 	&	$\Delta_{g}$~(rad/s)	&	 $\delta$ 	&	$\Pi_{\rm sig}$ (s)	&	 model and data	&	range	&	case	\\[1ex]
\hline																		
	\rule{0pt}{4ex}$314^{+14}_{-13}$	&	$58^{+40}_{-35}$	&	$ 0.00898^{+0.00014}_{-0.00007}$	&		&	$1.04^{+0.08}_{-0.11}$	&	2198	&	Step ; $l=1$	&	all	&	F	\\[2ex]
	$318^{+11}_{-11}$	&	$58^{+21}_{-32}$	&	$ 0.00783^{+0.00008}_{-0.00008}$	&		&	$2.42^{+0.10}_{-0.07}$	&	2521	&	Step ; $l=2$	&	all	&	G	\\[2ex]
	$334^{+17}_{-17}$	&	$0.060^{+0.037}_{-0.042}$	&	$0.0093^{+0.0006}_{-0.0004}$	&	$0.0014^{+0.0003}_{-0.0003}$	&	$0.91^{+0.35}_{-0.48}$	&	2122	&	Gaussian ; $l=1$ 	&	all	&	H	\\[2ex]
	$336^{+15}_{-19}$	&	$0.116^{+0.031}_{-0.089}$	&	$0.0082^{+0.0076}_{-0.0003}$	&	$0.0024^{+0.0013}_{-0.0009}$	&	$2.3^{+1.8}_{-0.6}$	&	2407	&	Gaussian ; $l=2$ 	&	all	&	I	\\[2ex]
\hline																														
 \end{tabular}
  }
\end{minipage}
\tablefoot{Parameters derived from the fits of a step-like glitch model (cases F and G) and a Gaussian-like glitch model (cases H and I) to the reduced period spacings of EPIC~211779126. The results on the left side of the table show the medians
and the 68~per cent confidence intervals of the inferred parameters, from left to right, the asymptotic reduced period spacing, the glitch amplitude, position and width, and the phase. The right side of the table indicates the periodicity of the glitch
signature and the choice of model and data. }
\end{table*}

%-------------------------------------------------------------

%-------------------------------------------------------------

\begin{table}[h!]
	\caption{Quality of the fits performed to the reduced period spacings of EPIC~211779126.}
	\label{tab:EPIC_chi2}
\begin{minipage}{0.4\textwidth}
 \centering
\resizebox{0.6\linewidth}{!}{%
  \begin{tabular}{l r c}
\hline					
\hline					
\rule{0pt}{3ex}	model and data	&	$\chi^2_{R40}$	&	case	\\[1ex]
\hline					
\rule{0pt}{4ex}Step ; $l=1$	&	1.36	&	F	\\[2ex]
Step ; $l=2$	&	0.88	&	G	\\[2ex]
Gaussian ; $l=1$ 	&	1.86	&	H	\\[2ex]
Gaussian ; $l=2$ 	&	1.28	&	I	\\[2ex]
\hline								
 \end{tabular}
  }
\end{minipage}
\tablefoot{The metric provided, $\chi^2_{R40}$, is a renormalized version of the reduced $\chi^2$, corresponding to a fixed reference value of the error, namely $\sigma^2 =(40~\rm{s})^2$ (see text for details).}
\end{table}

%-------------------------------------------------------------

\subsection{KIC~10001893}
\label{uzundag}
\begin{figure}
		\includegraphics[width=\columnwidth]{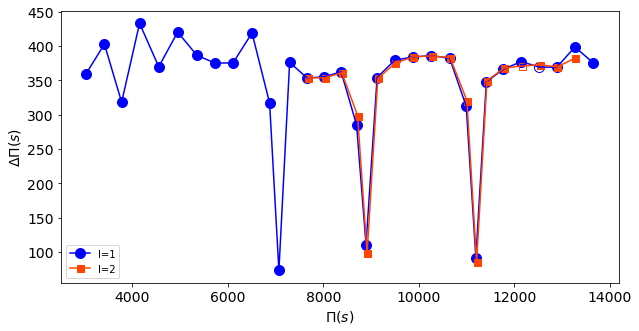}
      \includegraphics[width=\columnwidth]{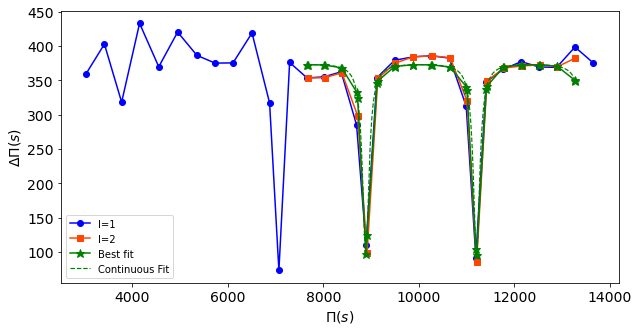}
		\includegraphics[width=\columnwidth]{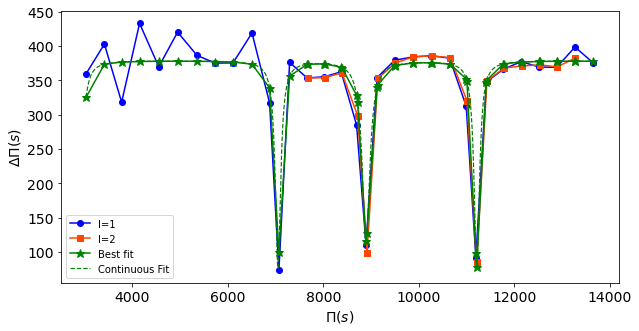}
    \caption{KIC~10001893 reduced period spacings as a function of the reduced period. Each reduced period spacing is plotted at the mid point between the reduced periods used in its computation. The parameters for the best fit solutions are given in Appendix~\ref{apC}.
    {Top panel:} observed reduced period spacings for $l=1$ (blue) and $l=2$ (red) modes. The discrete values are joined by straight lines to guide the eye. Symbols meaning as in the top panel of Fig.~\ref{fig:KIC1}.  {Middle panel:} the best fit solution (green) to the $l=1$  (blue) and $l=2$ (red) data in the range of periods where data on both mode degrees are available, for the case of a one step-like glitch (case J). The continuous analytical solution is represented by the dashed line, while the filled stars indicate the values of the solution at the reduced periods corresponding to the locations of the observations. {Bottom panel:} same as middle panel but for the case of two step-like glitches fitted to all data displayed in the panel (green) (case K).
}
    \label{fig:KIC2}
\end{figure}

The reduced period spacings of KIC~10001893 derived from the data of \cite{uzundag17} are shown in the top panel of Fig.~\ref{fig:KIC2}. Three dips are observed for the $l=1$ series of modes, but only two of these are covered by the $l=2$ modes. As pointed out by the authors, the fact that a dip is observed simultaneously for both mode degrees is particular important in a star like this, in which no multiplets are seen. In fact, they consider the significance of the first trapped mode less convincing and stress that although other unidentified modes could be associated with trapped modes, the absence of multiplets prevents them from having a clear preference. With this in mind, in our analysis we consider two cases, namely, a fit to the region where the $l=1$ and $l=2$ reduced periods overlap and a fit to all data available, covering the three dips. In both cases we follow the same strategy as before and do not attempt to fit the short-scale variations observed in the reduced period spacings, that in the present case appear at low values of the reduced periods. Therefore, for the first fit we adopt a model with a single step-like glitch. However, for the second fit a single glitch model is not adequate. When considering the three dips in the upper panel of Fig.~\ref{fig:KIC2}, we find that the distance between consecutive dips is too different to be adequately modelled by the impact of a single glitch, which predicts equally spaced dips in the reduced period spacings. Therefore, when fitting all $l=1$ and $l=2$ modes, we instead consider a model with two step-like glitches. These could hypothetically correspond to two different regions inside the star (\egc associated with the CO/He and the He/H transitions). 

The best-fit solutions obtained from the fits performed are shown in the middle panel (fit to the $l=1$ and $l=2$ overlap region) and bottom panel (fit to all data). Also, the median values of the model parameters and respective uncertainties are shown in Table~\ref{tab:KIC2_parameters}. Although both cases capture well the reduced period spacing dips in the fitted regions, the solutions are significantly different. In the first case (single glitch; overlap region) the glitch position is such as to reproduce the separation between the second and third period-spacing dips. In the second case (two glitches; all data) one of the glitches is positioned such as to reproduce the first and third dips, while the other glitch reproduces the single dip in the middle, without predicting any additional dip  within the observed range of reduced periods. Consequently, in this case both glitches are positioned at at buoyancy radius (or depth) smaller than the glitch inferred in the fit to the overlap region. 

%-------------------------------------------------------------
 \begin{table*}[h!]
	\caption{Seismic and glitch properties of KIC~10001893 derived from fits of a one step- and two step-like glitch models to the data.}
 \centering
\label{tab:KIC2_parameters}
\begin{minipage}{0.7\textwidth}
	\resizebox{\linewidth}{!}{%
  \begin{tabular}{l c c c |c l l c}
\hline																
\hline																
	\rule{0pt}{3ex}$\Delta\Pi_{\rm as}$ (s) 	&	 $A_{\rm st}$  or  $A_{\rm G}$  (rad/s)	&	$\tilde{\mathcal{W}}_{\rm g}^\star$ ( rad/s) 	&	 $\delta$ 	&	$\Pi_{\rm sig}$ (s)	&	 model and data	&	range	&	case	\\[1ex]
\hline																
	\rule{0pt}{4ex}$323.3^{+5.8}_{-6.4}$	&	$18.0^{+7.4}_{-4.2}$	&	$ 0.00849^{+0.00014}_{-0.00007}$	&	$2.88^{+0.14}_{-0.18}$	&	2325	&	1 Step ; $l=1,2$ 	&	common	&	J	\\[2ex]
	$334.5^{+8.3}_{-7.8}$	&	$48^{+36}_{-17}$	&	$ 0.00473$	&	$3.313^{+0.037}_{-0.040}$	&	4173	&	2 Step ; $l=1,2$ 	&	all	&	K	\\[2ex]
\hline																
 \end{tabular}
  }
\end{minipage}
\tablefoot{Parameters derived from the fits of a one step-like glitch model (case J) and two step-like glitches model (case K) to the reduced period spacings of KIC~10001893. The results on the left side of the table show the medians
and the 68~per cent confidence intervals of the inferred parameters, from left to right, the asymptotic reduced period spacing, the glitch amplitude and position, and the phase. The right side of the table indicates the periodicity of the glitch
signature and the choice of model and data.  For the fit with the two-glitch model, only the parameters for the first glitch are listed (see text for details).}
\end{table*}

%-------------------------------------------------------------

Inspection of the posterior distributions for the model parameters presented in Figs~\ref{fig:corner_KIC2_1} and \ref{fig:corner_KIC2_2} emphasises further differences between the two fits. In the one-glitch fit to the overlap region, all parameters are well constrained, including the glitch amplitude. The constraining power to the amplitude in this case results from a combination of factors. Firstly, the data from the $l=1$ and $l=2$ modes is in excellent agreement and focused on the region of the dips, so the constraints around the dips being fitted are stringer than in all cases considered before. Secondly, the one-glitch model fits the data significantly better than in previous cases, as seen from the error distribution in Fig~\ref{fig:corner_KIC2_1} (whose median value is $\sim 19$~s, compared to median values $\sim 45-60$~s in previous cases). In the two-glitch case, Fig.~\ref{fig:corner_KIC2_2} shows the difficulty in properly constraining the second glitch (the one labelled `out' that reproduces the second period-spacing dip). This is expected, as all glitch positions closer to one of the edges of the buoyancy cavity than a given value are possible solutions. In this case, that value corresponds to the glitch position that would induce a signal with dips separated by as much as the range of reduced periods considered. This threshold induces the abrupt right-hand-side drop in the posterior distribution of $\omega_{\rm g, out}^\star$ for the second glitch, which we arbitrarily take to be the one in the outer cavity (\cf Fig.~\ref{fig:corner_KIC2_2}), to the left of which we observe a plateau of similarly good solutions. 

% Example figure

\section{Discussion}
\label{sec:discussion}

\subsection{Glitch nature}

The glitch properties inferred for the three stars in this study indicate that the same phenomenon is being observed in all cases. This is illustrated in Fig. \ref{fig:sdb_clump} where we can see the similarity of the glitch positions and glitch amplitudes (albeit the latter inferred with large uncertainties). The figure shows also the asymptotic reduced period spacings for the three stars, confirming the overall similarity of their radiative inner structure. The cases illustrated for each star are identified on the x-axis of the figure.  For KIC10553698A (KIC$_1$ in the figure), we show the fits to the reduced periods up to 111000~s, the most reliable data range, as discussed in Sect.~\ref{sec:kic1}. Fitting all reduced periods listed in Tables~\ref{tab:KIC1_all} leads to only a slightly smaller value of the glitch position. For KIC10001893 (KIC$_2$), we show only the results from the fit with a one-glitch model, for consistency with the analyses shown for the other stars. The fit of the data assuming a two-glitch model leads to a very different solution, where neither glitch is located in a position comparable to that seen in the other stars. This reinforces the idea that the one-glitch model is likely more adequate to explain the data for this star, hence, that the characteristic periodicity of the glitch signature in KIC10001893 is given by the distance between two consecutive dips in the stars' reduced period spacings. In this interpretation, the uneven separation between the three dips observed in the period spacings results  either from a misidentification of the trapped mode in the oscillation spectrum, or potentially from the glitch shape being more complex than the adopted step function. In relation to the former possibility, \cite{uzundag17} point out that the significance of the first trapped mode they identify for $l=1$, between $n=17$ and $n=18$, is less convincing than that of the second trapped mode, and that there are other unidentified peaks that could be associated to trapped modes but the lack of multiplets makes it difficult for the authors to make a definitive selection.  

{The fact that the period-spacing dips in two of the three stars (KIC10001893 and KIC10553698A) do not significantly depend on the reduced periods, indicates that the glitch is sharp.} In the third star (EPIC211779126) only two dips are observed, followed by a long plateau in which the reduced period spacings vary only with small amplitude. This could be an indication of a wider structural variation which would cease to act as a glitch for the modes of longer period (shorter characteristic scale). However, comparison between fits to the data of this star with step-like and Gaussian-like glitch models do not necessarily corroborate that interpretation. In fact, the fits for the two cases have similar qualities (Table~\ref{tab:EPIC_chi2}) and inspection of the middle and bottom panels of Fig.~\ref{fig:EPIC} seems to indicate that the step-like glitch provides a better representation of the flat top of the period spacings. Together with the fact that the glitch position inferred from the fits to the data of EPIC211779126 is similar to that inferred for the other two stars, we argue that it is more likely that the period-spacing signature in this star is produced by the same structural transition as in the other two stars. 

{When considering the nature of the structural transition that is causing the signature observed in the reduced period spacings, it is important to recall that the thermal and chemical stratifications of the layers adjacent to the convectively-unstable core of sdB stars are still largely unknown. In this context, the interpretation of the glitch position needs to bear in mind that g modes do not propagate into the adiabatically stratified part of the core. Thus, the glitch location inferred from our fits is blind to those layers, measuring only the buoyancy distance from the edge of the convective core, \iec  the border of the adiabatically stratified region as defined in Sect.~\ref{sec:ps}. Recently, \cite{guyot2025} have studied the pulsation spectra of sdB models with different core properties illustrating the sensitive of the period spacing modulation to the stratification of the core. Comparing the values of $\Pi_{\rm sig}$ that we infer with the results from their models, it seems that the glitches in these stars are best compatible with the C-O/He transition in sdB stars with a low core-He abundance and a relatively small core mass. In addition, the data analysed here potentially hints at a slightly sharper glitch than the transition in their models, as their results show a smooth decrease in the depth of the period-spacing dips with increasing period. This typical characteristic of the signatures from non-sharp transitions is not evident in the observational data. However, with only three period-spacing dips it is impossible to establish the significance of this difference between the authors' models and the observational data.  } 

\subsection{Comparison with glitches observed in helium core-burning red giants}
\begin{figure*}[h!]
    \includegraphics[width=2\columnwidth]{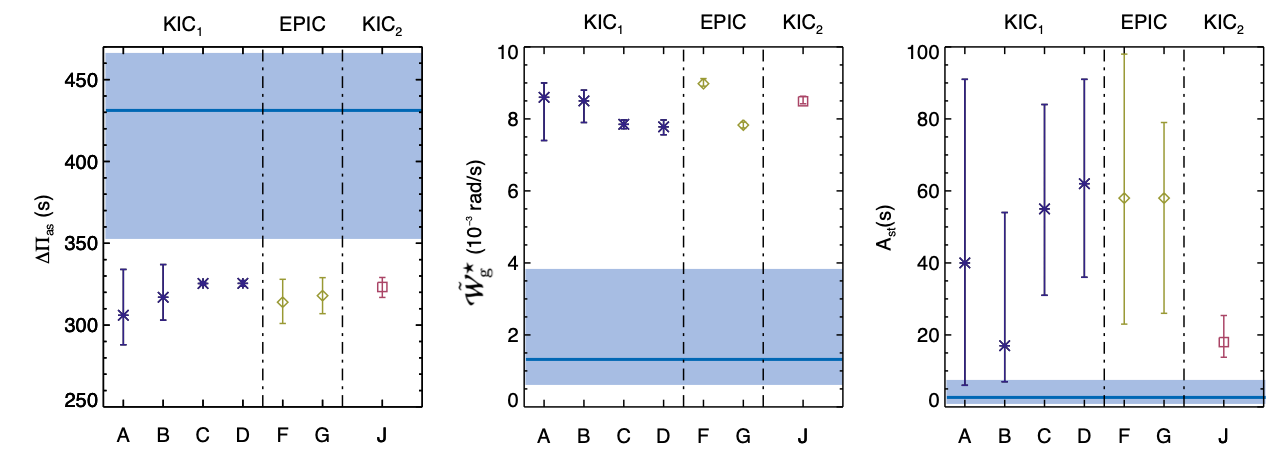}
   \caption{Physical parameters inferred from the fits with a one step-like glitch model. The stars are identified in the top axis as KIC$_1$=KIC10553698A, EPIC=EPIC211779126, and KIC$_2$=KIC10001893 and the respective results are separated by a dotted-dashed line. The labels in the x-axis identify the case according to the results given in Tables~\ref{tab:KIC1_parameters}, \ref{tab:EPIC_parameters}, and \ref{tab:KIC2_parameters}. From left to right, the panels show, with symbols and error bars, the median and respective 68~per cent uncertainties, on the asymptotic reduced period spacing, degree-independent glitch buoyancy radius, and glitch amplitude, respectively. The blue line and shaded blue area in each panel show, respectively, the median and range of values found by \cite{vrard22} for the same parameters in the study of a sample of helium core-burning red giant stars. }
    \label{fig:sdb_clump}
\end{figure*}

Subdwarf B stars are likely the product of non-standard evolution involving the stripping of the outer layers of a star in the red giant branch (RGB). Therefore, it is important to investigate how the cores of helium core-burning red giant stars, which arise from standard stellar evolution, compare to those of sdB stars, which reach helium core burning through non-standard evolutionary channels. The inference of the core glitch properties in these stars provides the opportunity to address this question directly from the analysis of the data, \iec in a way which is independent from our ability to build stellar evolution or static stellar models. With this in mind, in Fig.~\ref{fig:sdb_clump} we overplot in blue the range of asymptotic reduced period spacings, glitch locations, and glitch amplitudes derived by \cite{vrard22} for 23 helium core-burning red giants. The blue horizontal line marks the median of the respective quantities which were converted to their reduced form when relevant.

The difference in the results for the two samples is striking. In the case of the asymptotic reduced period spacings, it reflects an overall difference in the radiative interior of the two types of stars. In the case of the glitch amplitude, it could reflect a difference in the relative amplitude of the discontinuity in the buoyancy frequency at the chemical transition, or simply result from the finite width of the transition. In fact, the periods of the modes observed in the sdB stars are shorter, by a factor of $\sim 5$, than the periods of mixed modes observed in helium core-burning red giants. Thus, a slow dependence of the amplitude of the glitch signature on the reduced periods, resulting from a hypothetical intrinsic width of the glitch, would lead to a much smaller amplitude of the glitch signature at the longer periods observed in the helium core-burning red giants than in the range of periods observed in sdB stars. Since both the helium core-burning red giant data and the sdB data were fitted with step-like glitch models that do not incorporate a period dependence of the glitch-signature amplitude, the hypothetical glitch width would be wrongly translated into a difference in the amplitudes inferred for the structural variations in the two types of stars. 

However, perhaps the most interesting difference observed in the properties inferred for the sdB and the helium core-burning red giants concerns the glitch position. Due to its strict association to the periodicity of the glitch signature, this result is robust and cannot be explained by a change in the shape of the glitch. If the signature observed in the data of sdB and helium core-burning red giant stars is indeed associated with a transition close to the convective core, then the glitch position given in terms of the absolute quantity \wgstarlt\, provides a direct measurement of the integral of $N/r$ from the {edge of the convective core}, where presumably the inner turning point of the modes is located, to the maximum extent of the chemical mixing into the radiative core (\cf\, Eq.~(\ref{eq:bradius})). According to our results, that extent is significantly different in the two types of stars.

\section{Conclusions}
\label{sec:conclusions}

In this work we used an analytical model developed in the works of \cite{cunha19,cunha24} for g modes to directly infer information on structural glitches inside three sdB stars, without recourse to specific stellar  models. Our main conclusions are as follows.
\begin{itemize}[label=$-$]
    \item Despite their simplicity, in all three stars the analytical models capture well the most significant variations of the periods around the equally-spaced asymptotic prediction;
    \item According to the model inferences, the structural glitches inducing the observed period perturbations have similar properties in all three stars. They are consistent with sharp discontinuities located close to one of the turning points of the g-mode cavity, likely the C-O/He transition. {Comparison with recent results from \cite{guyot2025} favour a scenario where these are evolved sdB stars, with low core-He abundance and a relatively small core mass};
    \item  The results indicate that the glitch properties in sdB stars are systematically different from those of glitches in helium core-burning red giants derived by \cite{vrard22}.  In particular, the sdB glitches are located at significantly larger buoyancy radii, indicating {either a more extensive mixing region beyond the edge of the convective core  or a different stratification of that mixing region. }  
\end{itemize}

Whether or not the differences found between the glitches in sdB and helium core-burning red giant stars can be explained by the adjustment of the stellar structure to the stripping of the outer layers is key to test and understand the evolution channels leading to the formation of sdB stars. That is best answered by simulations of the stellar structure and shall be addressed in future work. 

\begin{acknowledgements}
This work has been supported by Fundação para a Ciência e Tecnologia FCT-MCTES, Portugal, through national funds by these grants UIDB/04434/2020 (DOI: 10.54499/UIDB/04434/2020), UIDP/04434/2020.FCT (DOI: 10.54499/UIDP/04434/2020) and 2022.03993.PTDC (DOI:10.54499/2022.03993.PTDC). MSC is funded by FCT-MCTES by the contract with reference 2023.09303.CEECIND/CP2839/CT0003 (DOI:10.54499/2023.09303.CEECIND/CP2839/CT0003). AF acknowledges the funding from FEDER (through COMPETE2020 programme) and FCT under the project 'BreakStarS', attributed by CAUP with reference: CIAAUP-06/2022-BII.

\end{acknowledgements}

%%%%%%%%%%%%%%%%%%%%%%%%%%%%%%%%%%%%%%%%%%%%%%%%%%

%%%%%%%%%%%%%%%%%%%% REFERENCES %%%%%%%%%%%%%%%%%%

% The best way to enter references is to use BibTeX:

\bibliographystyle{aa}
\bibliography{solar-like} % if your bibtex file is called example.bib

% Alternatively you could enter them by hand, like this:
% This method is tedious and prone to error if you have lots of references
%\begin{thebibliography}{99}
%\bibitem[\protect\citeauthoryear{Author}{2012}]{Author2012}
%Author A.~N., 2013, Journal of Improbable Astronomy, 1, 1
%\bibitem[\protect\citeauthoryear{Others}{2013}]{Others2013}
%Others S., 2012, Journal of Interesting Stuff, 17, 198
%\end{thebibliography}

%%%%%%%%%%%%%%%%%%%%%%%%%%%%%%%%%%%%%%%%%%%%%%%%%%

%%%%%%%%%%%%%%%%% APPENDICES %%%%%%%%%%%%%%%%%%%%%

\appendix
\begin{appendix}
\onecolumn
\section{Data}
%\red{add tables with data}
\label{apA}

\begin{table*}[!ht]
	\centering
	\caption{KIC~10553698A: observed periods used for cases A to D of Table~\ref{tab:KIC1_parameters} \cite[source][]{ostensen14}. 
    }
	\label{tab:KIC1}
	\begin{tabular}{lcccclccc} % four columns, alignment for each
\hline																
$l$	&	$n$	&	$P$ (s)	&	$\Pi$ (s)	&	&	$l$	&	$n$	&	$P$ (s)	&	$\Pi$ (s)	\\
\hline																
1	&	9	&	2518.00	&	3560.99	&	&	--	&	--	&	--	&	--	\\
1	&	10	&	2759.12	&	3901.98	&	&	--	&	--	&	--	&	--	\\
1	&	11	&	3036.15	&	4293.76	&	&	--	&	--	&	--	&	--	\\
1	&	12	&	3310.43	&	4681.66	&	&	--	&	--	&	--	&	--	\\
1	&	t 	&	3482.01	&	4924.31	&	&	--	&	--	&	--	&	--	\\
1	&	13	&	3611.60	&	5107.57	&	&	--	&	--	&	--	&	--	\\
1	&	14	&	3861.40	&	5460.84	&	&	2	&	14	&	2235.60	&	5476.08	\\
1	&	15	&	4116.26	&	5821.27	&	&	2	&	15	&	2384.42	&	5840.61	\\
1	&	16	&	4392.22	&	6211.54	&	&	2	&	16	&	2543.47*	&	6230.20	\\
1	&	17	&	4672.27	&	6607.59	&	&	2	&	17	&	2701.68	&	6617.74	\\
1	&	18	&	4947.10	&	6996.26	&	&	2	&	18	&	2864.21	&	7015.85	\\
1	&	19	&	5200.90*	&	7355.18	&	&	2	&	19	&	3002.42$^+$	&	7354.40$^+$	\\
1	&	t 	&	5413.48	&	7655.82	&	&	2	&	t	&	3123.99*	&	7652.18	\\
1	&	20	&	5469.33	&	7734.80	&	&	2	&	20	&	3157.78	&	7734.95	\\
1	&	21	&	5698.87	&	8059.42	&	&	2	&	21	&	3295.35	&	8071.93	\\
1	&	22	&	5957.62	&	8425.35	&	&	2	&	22	&	3453.78	&	8460.00	\\
1	&	23	&	6219.27	&	8795.38	&	&	2	&	23	&	3596.00*	&	8808.37	\\
1	&	24	&	6469.42	&	9149.14	&	&	2	&	24	&	3738.30*	&	9156.93	\\
1	&	25	&	6735.73	&	9525.76	&	&	2	&	25	&	3897.85*	&	9547.74	\\
1	&	26	&	7012.15	&	9916.68	&	&	2	&	26	&	4057.45	&	9938.68	\\
1	&	t 	&	7116.32	&	10064.00	&	&	2	&	t 	&	4116.26$^\dagger$	&	10082.74	\\
1	&	27	&	7278.81	&	10293.79	&	&	2	&	27	&	4216.30	&	10327.78	\\
1	&	28	&	7560.15	&	10691.67	&	&	2	&	28	&	4370.81	&	10706.25	\\
1	&	29	&	7811.72*	&	11047.44	&	&	2	&	29	&	4519.28	&	11069.93	\\
1	&	30	&	8081.92	&	11429.56	&	&	2	&	30	&	4658.74	&	11411.54	\\
1	&	31	&	8333.06$^+$	&	11784.73$^+$	&	&	2	&	31	&	4810.36	&	11782.93	\\
1	&	32	&	8598.56$^+$	&	12160.20$^+$	&	&	2	&	32	&	4949.51$^+$	&	12123.77$^+$	\\
1	&	33	&	8847.47*	&	12512.21	&	&	2	&	33	&	5118.15$^+$	&	12536.86$^+$	\\
1	&	34	&	9112.01	&	12886.33	&	&	2	&	34	&	5239.75	&	12834.71	\\
1	&	35	&	9397.62	&	13290.24	&	&	2	&	35	&	5394.41	&	13213.55	\\
1	&	36	&	9588.62	&	13560.36	&	&	-- &	--	&	--	&	--	\\
\hline																
 \multicolumn{9}{l}{$*$ average of multiplet components in \cite{ostensen14}.} \\
 \multicolumn{9}{l}{$+$ estimated from figure 10 of \cite{ostensen14}.} \\
\multicolumn{9}{l}{$^{\dagger}$ $f_{63}$ in \cite{ostensen14}, as per discussion in that paper.}\\
	\end{tabular}
    \tablefoot{The first four columns show the degree, asymptotic radial order, period and reduced period for the $l=1$ modes and the last four columns show the same information for the $l=2$ modes. Trapped modes are identified with a "t" in the radial order column. }
\end{table*}

\begin{table*}[!ht]
	\centering
	\caption{KIC~10553698A: observed periods used for case E identified in Table~\ref{tab:KIC1_parameters} in this work \cite[source][]{ostensen14}.}
	\label{tab:KIC1_all}
	\begin{tabular}{lcccclccc} % four columns, alignment for each
\hline																
$l$	&	$n$	&	$P$ (s)	&	$\Pi$ (s)	&	&	$l$	&	$n$	&	$P$ (s)	&	$\Pi$ (s)	\\
\hline																
1	&	9	&	2518.00	&	3560.99	&	&	--	&	--	&	--	&	--	\\
1	&	10	&	2759.12	&	3901.98	&	&	--	&	--	&	--	&	--	\\
1	&	11	&	3033.71	&	4290.31	&	&	--	&	--	&	--	&	--	\\
1	&	11	&	3036.15	&	4293.76	&	&	--	&	--	&	--	&	--	\\
1	&	12	&	3307.53	&	4677.55	&	&	--	&	--	&	--	&	--	\\
1	&	12	&	3308.96	&	4679.58	&	&	--	&	--	&	--	&	--	\\
1	&	12	&	3310.43	&	4681.66	&	&	--	&	--	&	--	&	--	\\
1	&	t	&	3479.94	&	4921.38	&	&	--	&	--	&	--	&	--	\\
1	&	t	&	3482.01	&	4924.31	&	&	--	&	--	&	--	&	--	\\
1	&	13	&	3609.77	&	5104.99	&	&	--	&	--	&	--	&	--	\\
1	&	13	&	3611.60	&	5107.57	&	&	--	&	--	&	--	&	--	\\
1	&	13	&	3613.45	&	5110.19	&	&	--	&	--	&	--	&	--	\\
1	&	14	&	3859.04	&	5457.51	&	&	2	&	14	&	2235.60	&	5476.08	\\
1	&	14	&	3861.40	&	5460.84	&	&	2	&	14	&	2238.19	&	5482.42	\\
1	&	14	&	3863.78	&	5464.21	&	&	2	&	14	&	2239.49	&	5485.61	\\
1	&	15	&	4116.26	&	5821.27	&	&	2	&	15	&	2383.02	&	5837.18	\\
1	&	15	&	4122.81	&	5830.53	&	&	2	&	15	&	2384.42	&	5840.61	\\
--	&	--	&	--	&	--	&	&	2	&	15	&	2388.65	&	5850.97	\\
1	&	16	&	4392.22	&	6211.54	&	&	2	&	16	&	2540.44	&	6222.78	\\
--	&	--	&	--	&	--	&	&	2	&	16	&	2541.96	&	6226.50	\\
--	&	--	&	--	&	--	&	&	2	&	16	&	2544.98	&	6233.90	\\
--	&	--	&	--	&	--	&	&	2	&	16	&	2546.50	&	6237.63	\\
1	&	17	&	4672.27	&	6607.59	&	&	2	&	17	&	2701.68	&	6617.74	\\
1	&	17	&	4675.28	&	6611.84	&	&	--	&	--	&	--	&	--	\\
1	&	18	&	4943.02	&	6990.49	&	&	2	&	18	&	2864.21	&	7015.85	\\
1	&	18	&	4947.10	&	6996.26	&	&	2	&	18	&	2866.33	&	7021.05	\\
1	&	18	&	4950.90	&	7001.63	&	&	--	&	--	&	--	&	--	\\
1	&	19	&	5193.95	&	7345.35	&	&	--	&	--	&	--	&	--	\\
1	&	19	&	5197.76	&	7350.74	&	&	--	&	--	&	--	&	--	\\
1	&	19	&	5202.50	&	7357.45	&	&	--	&	--	&	--	&	--	\\
1	&	19	&	5209.37	&	7367.16	&	&	--	&	--	&	--	&	--	\\
1	&	t	&	5413.48	&	7655.82	&	&	2	&	t	&	3123.22	&	7650.30	\\
--	&	--	&	--	&	--	&	&	2	&	t	&	3124.75	&	7654.04	\\
1	&	20	&	5464.97	&	7728.63	&	&	2	&	20	&	3157.78	&	7734.95	\\
1	&	20	&	5469.33	&	7734.80	&	&	2	&	20	&	3160.05	&	7740.51	\\
1	&	21	&	5698.87	&	8059.42	&	&	2	&	21	&	3295.35	&	8071.93	\\
1	&	21	&	5703.21	&	8065.56	&	&	--	&	--	&	--	&	--	\\
1	&	21	&	5710.35	&	8075.65	&	&	--	&	--	&	--	&	--	\\
1	&	22	&	5957.62	&	8425.35	&	&	2	&	22	&	3450.15	&	8451.11	\\
1	&	22	&	5960.53	&	8429.46	&	&	2	&	22	&	3453.78	&	8460.00	\\
1	&	22	&	5964.69	&	8435.35	&	&	--	&	--	&	--	&	--	\\
1	&	22	&	5969.05	&	8441.51	&	&	--	&	--	&	--	&	--	\\
1	&	22	&	5972.27	&	8446.07	&	&	--	&	--	&	--	&	--	\\
1	&	23	&	6219.27	&	8795.38	&	&	2	&	23	&	3588.54	&	8790.09	\\
1	&	23	&	6225.69	&	8804.46	&	&	2	&	23	&	3603.45	&	8826.61	\\
1	&	23	&	6230.39	&	8811.10	&	&	--	&	--	&	--	&	--	\\
--	&	--	&	--	&	--	&	&	2	&	24	&	3736.51	&	9152.54	\\
--	&	--	&	--	&	--	&	&	2	&	24	&	3740.09	&	9161.31	\\
\hline			
\end{tabular}
\end{table*}

%-------------- continuation---------

 \begin{table*}[ht]
	\centering
    \ContinuedFloat
	\caption{KIC~10553698A: continuing from previous page}
	\label{tab:KIC1_all}
	\begin{tabular}{lcccclccc} % four columns, alignment for each
\hline																
$l$	&	$n$	&	$P$ (s)	&	$\Pi$ (s)	&	&	$l$	&	$n$	&	$P$ (s)	&	$\Pi$ (s)	\\
\hline																
1	&	25	&	6735.73	&	9525.76	&	&	2	&	25	&	3893.94	&	9538.17	\\
1	&	25	&	6738.75	&	9530.03	&	&	2	&	25	&	3901.76	&	9557.32	\\
1	&	25	&	6743.41	&	9536.62	&	&	--	&	--	&	--	&	--	\\
1	&	26	&	7004.67	&	9906.10	&	&	2	&	26	&	4057.45	&	9938.68	\\
1	&	26	&	7012.15	&	9916.68	&	&	--	&	--	&	--	&	--	\\
1	&	26	&	7020.98	&	9929.17	&	&	--	&	--	&	--	&	--	\\
1	&	t	&	7116.32	&	10064.00	&	&	--	&	--	&	--	&	--	\\
1	&	27	&	7271.22	&	10283.06	&	&	2	&	27	&	4216.30	&	10327.78	\\
1	&	27	&	7274.34	&	10287.47	&	&	--	&	--	&	--	&	--	\\
1	&	27	&	7278.81	&	10293.79	&	&	--	&	--	&	--	&	--	\\
1	&	27	&	7286.29	&	10304.37	&	&	--	&	--	&	--	&	--	\\
1	&	28	&	7560.15	&	10691.67	&	&	2	&	28	&	4366.62	&	10695.99	\\
--	&	--	&	--	&	--	&	&	2	&	28	&	4370.81	&	10706.25	\\
--	&	--	&	--	&	--	&	&	2	&	28	&	4378.71	&	10725.61	\\
1	&	29	&	7800.71	&	11031.87	&	&	2	&	29	&	4519.28	&	11069.93	\\
1	&	29	&	7809.28	&	11043.99	&	&	--	&	--	&	--	&	--	\\
1	&	29	&	7814.84	&	11051.85	&	&	--	&	--	&	--	&	--	\\
1	&	29	&	7822.05	&	11062.05	&	&	--	&	--	&	--	&	--	\\
--	&	--	&	--	&	--	&	&	2	&	30	&	4658.74	&	11411.54	\\
--	&	--	&	--	&	--	&	&	2	&	31	&	4797.81	&	11752.19	\\
--	&	--	&	--	&	--	&	&	2	&	31	&	4805.24	&	11770.39	\\
--	&	--	&	--	&	--	&	&	2	&	31	&	4810.36	&	11782.93	\\
--	&	--	&	--	&	--	&	&	2	&	31	&	4815.86	&	11796.40	\\
--	&	--	&	--	&	--	&	&	2	&	31	&	4820.98	&	11808.94	\\
1	&	33	&	8845.20	&	12509.00	&	&	--	&	--	&	--	&	--	\\
1	&	33	&	8849.74	&	12515.42	&	&	--	&	--	&	--	&	--	\\
1	&	34	&	9112.01	&	12886.33	&	&	2	&	34	&	5239.75	&	12834.71	\\
1	&	35	&	9390.67	&	13280.41	&	&	2	&	35	&	5376.51	&	13169.71	\\
1	&	35	&	9397.62	&	13290.24	&	&	2	&	35	&	5385.58	&	13191.92	\\
1	&	35	&	9403.64	&	13298.76	&	&	2	&	35	&	5394.41	&	13213.55	\\
1	&	36	&	9588.62	&	13560.36	&	&	--	&	--	&	--	&	--	\\
\hline			
\end{tabular}
\tablefoot{ The first four columns show the degree, asymptotic radial order, period and reduced period for the $l=1$ modes and the last four columns show the same information for the $l=2$ modes. Trapped modes are identified with a "t" in the radial order column.}
\end{table*}

\begin{table*}
	\centering
	\caption{EPIC~211779126: observed periods used in this work \cite[source][]{baran17}. 
    }
	\label{tab:EPIC}
	\begin{tabular}{lcccclccc} % four columns, alignment for each
		\hline
$l$	&	$n$	&	$P$ (s)	&	$\Pi$ (s)	&	&	$l$	&	$n$	&	$P$ (s)	&	$\Pi$ (s)	\\
\hline																
1	&	11	&	2953.16	&	4176.40	&	&	--	&	--	&	--	&	--	\\
1	&	12	&	3212.44	&	4543.08	&	&	2	&	12	&	1846.31	&	4522.52	\\
1	&	13	&	3483.71	&	4926.71	&	&	2	&	13	&	2017.55	&	4941.97	\\
1	&	t	&	3738.32	&	5286.78	&	&	2	&	t	&	2051.58	&	5025.32	\\
1	&	14	&	3752.77	&	5307.22	&	&	2	&	14	&	2173.25	&	5323.35	\\
1	&	15	&	4014.13	&	5676.84	&	&	2	&	15	&	2322.50$^+$	&	5688.94$^+$	\\
1	&	16	&	4273.69	&	6043.91	&	&	2	&	16	&	2473.41	&	6058.59	\\
1	&	17	&	4533.50	&	6411.34	&	&	2	&	17	&	2623.78	&	6426.92	\\
1	&	18	&	4800.54	&	6788.99	&	&	2	&	18	&	2776.00$^+$	&	6799.78$^+$	\\
1	&	19	&	5058.68	&	7154.05	&	&	2	&	19	&	2927.14	&	7170.00	\\
1	&	20	&	5287.65	&	7477.87	&	&	2	&	20	&	3053.81	&	7480.28	\\
1	&	t	&	5347.59	&	7562.63	&	&	2	&	t	&	3091.29	&	7572.08	\\
1	&	21	&	5551.85	&	7851.50	&	&	2	&	21	&	3220.92	&	7889.61	\\
1	&	22	&	5777.68	&	8170.87	&	&	2	&	22	&	3349.52	&	8204.61	\\
1	&	23	&	6042.66	&	8545.61	&	&	2	&	23	&	3490.52	&	8549.99	\\
1	&	24	&	6292.08	&	8898.34	&	&	2	&	24	&	3638.48	&	8912.42	\\
1	&	25	&	6565.99	&	9285.71	&	&	2	&	25	&	3793.91	&	9293.14	\\
1	&	26	&	6835.27	&	9666.53	&	&	2	&	26	&	3939.96	&	9650.89	\\
1	&	27	&	7075.14	&	10005.76	&	&	2	&	27	&	4091.32	&	10021.65	\\
1	&	28	&	7337.83	&	10377.26	&	&	2	&	28	&	4238.37	&	10381.84	\\
1	&	29	&	7616.73	&	10771.68	&	&	2	&	29	&	4391.16	&	10756.10	\\
1	&	30	&	7841.29	&	11089.26	&	&	2	&	30	&	4517.12	&	11064.64	\\
1	&	31	&	8054.77	&	11391.16	&	&	2	&	31	&	4673.12	&	11446.76	\\
1	&	32	&	8340.98	&	11795.93	&	&	2	&	32	&	4822.30	&	11812.17	\\
1	&	33	&	8603.63	&	12167.37	&	&	2	&	33	&	4968.20	&	12169.55	\\
1	&	34	&	8849.56	&	12515.17	&	&	2	&	34	&	5104.91	&	12504.42	\\
1	&	35	&	9131.59	&	12914.02	&	&	2	&	35	&	5275.93	&	12923.34	\\
1	&	36	&	9355.41	&	13230.55	&	&	2	&	36	&	5419.17	&	13274.20	\\
1	&	37	&	9638.55	&	13630.97	&	&	2	&	37	&	5574.45	&	13654.56	\\
1	&	38	&	9912.77	&	14018.77	&	&	2	&	38	&	5729.02	&	14033.18	\\
1	&	39	&	10147.13	&	14350.21	&	&	2	&	39	&	5867.17	&	14371.57	\\
1	&	40	&	10364.84	&	14658.10	&	&	2	&	40	&	5987.67	&	14666.74	\\
1	&	41	&	10617.97	&	15016.08	&	&	2	&	41	&	6165.99	&	15103.53	\\
1	&	42	&	10985.39	&	15535.69	&	&	2	&	42	&	6327.11	&	15498.19	\\
1	&	43	&	11179.43	&	15810.10	&	&	2	&	43	&	6453.28	&	15807.24	\\
\hline																		
 \multicolumn{9}{l}{$+$ estimated from figure 10 of \cite{baran17}.} \\
	\end{tabular}
    \tablefoot{ The first four columns show the degree, asymptotic radial order, period, and reduced period for the $l=1$ modes and the last four columns show the same information for the $l=2$ modes. Trapped modes are identified with a "t" in the radial order column. }
\end{table*}

\begin{table*}
	\centering
	\caption{KIC~10001893: observed periods used in this work \cite[source][]{uzundag17}.  
    }
	\label{tab:KIC2}
	\begin{tabular}{lcccclccc} % four columns, alignment for each
\hline																
$l$	&	$n$	&	$P$ (s)	&	$\Pi$ (s)	&	&	$l$	&	$n$	&	$P$ (s)	&	$\Pi$ (s)	\\
\hline																
1	&	7	&	2015.5	&	2850.3	&	&	--	&	--	&	--	&	--	\\
1	&	8	&	2270.1	&	3210.4	&	&	--	&	--	&	--	&	--	\\
1	&	9	&	2555.1	&	3613.5	&	&	--	&	--	&	--	&	--	\\
\hline	&	10	&	2780.2	&	3931.8	&	&	--	&	--	&	--	&	--	\\
1	&	11	&	3086.6	&	4365.1	&	&	--	&	--	&	--	&	--	\\
1	&	12	&	3348.2	&	4735.1	&	&	--	&	--	&	--	&	--	\\
1	&	13	&	3645.6	&	5155.7	&	&	--	&	--	&	--	&	--	\\
1	&	14	&	3919.0	&	5542.3	&	&	--	&	--	&	--	&	--	\\
1	&	15	&	4184.4	&	5917.6	&	&	--	&	--	&	--	&	--	\\
1	&	16	&	4450.0	&	6293.3	&	&	--	&	--	&	--	&	--	\\
1	&	17	&	4746.5	&	6712.6	&	&	--	&	--	&	--	&	--	\\
1	&	t	&	4970.8	&	7029.8	&	&	--	&	--	&	--	&	--	\\
1	&	18	&	5022.3	&	7102.6	&	&	--	&	--	&	--	&	--	\\
1	&	19	&	5288.4	&	7478.9	&	&	2	&	19	&	3060.1	&	7495.7	\\
1	&	20	&	5538.5	&	7832.6	&	&	2	&	20	&	3204.5	&	7849.4	\\
1	&	21	&	5789.7	&	8187.9	&	&	2	&	21	&	3349.0	&	8203.3	\\
1	&	22	&	6046.1	&	8550.5	&	&	2	&	22	&	3496.5	&	8564.6	\\
1	&	t	&	6247.7	&	8835.6	&	&	2	&	t	&	3618.5	&	8863.5	\\
1	&	23	&	6325.8	&	8946.0	&	&	2	&	23	&	3658.5	&	8961.5	\\
1	&	24	&	6576.3	&	9300.3	&	&	2	&	24	&	3802.6	&	9314.4	\\
1	&	25	&	6844.6	&	9679.7	&	&	2	&	25	&	3955.8	&	9689.7	\\
1	&	26	&	7116.4	&	10064.1	&	&	2	&	26	&	4112.8	&	10074.3	\\
1	&	27	&	7389.4	&	10450.2	&	&	2	&	27	&	4270.3	&	10460.1	\\
1	&	28	&	7659.9	&	10832.7	&	&	2	&	28	&	4426.8	&	10843.4	\\
1	&	t	&	7880.8	&	11145.1	&	&	2	&	t	&	4557.7	&	11164.0	\\
1	&	29	&	7945.6	&	11236.8	&	&	2	&	29	&	4592.3	&	11248.8	\\
1	&	30	&	8191.6	&	11584.7	&	&	2	&	30	&	4735.1	&	11598.6	\\
1	&	31	&	8450.7	&	11951.1	&	&	2	&	31	&	4885.3	&	11966.5	\\
1	&	32	&	8717.6	&	12328.5	&	&	2	&	32	&	5040.9$^+$	&	12347.7$^+$	\\
1	&	33	&	 8979.7$^+$	&	12699.2$^+$	&	&	2	&	33	&	5189.0	&	12710.4	\\
1	&	34	&	9243.0	&	13071.6	&	&	2	&	34	&	5340.0	&	13080.3	\\
1	&	35	&	9522.0	&	13466.1	&	&	2	&	35	&	5496.2	&	13462.9	\\
1	&	36	&	9787.4	&	13841.5	&	&	--	&	--	&	--	&	--	\\
\hline																
 \multicolumn{9}{l}{$+$ estimated from figure 6 of \cite{uzundag17}. See also \cite{Reed11}.} \\
	\end{tabular}
    \tablefoot{The first four columns show the degree, asymptotic radial order, period, and reduced period for the $l=1$ modes and the last four columns show the same information for the $l=2$ modes. Trapped modes are identified with a "t" in the radial order column.}
\end{table*}

\FloatBarrier
\section{Two-glitch model}
\label{apB}
Here we expand on the analysis presented in \cite{cunha19} to show how the eigenvalue condition derived for the one glitch model can be expanded to consider the case of two glitches, so far as they are sufficiently apart that the eigenfunction in between the glitches can be assumed to be well represented by the asymptotic solution. We take, as an example, the case of two step-like glitches, one on each side of the g-mode cavity. Other cases (\egc for glitches with different shapes) can be derived following similar steps. Without loss of generality, we shall assume that $N$ drops abruptly in the inner cavity and then increases abruptly, in the outer cavity. As discussed by \cite{cunha24}, reverting the sign of these structural changes does not change the glitch-induced signature in the periods, so far as the inferred amplitude is restricted to be positive and its value is interpreted as  $A_{\rm st}^{\pm}=[N_{\rm +}/N_{\rm -}]_{r^\star}-1$, where the subscripts $+$ and $-$ indicate the largest and the smallest of the two values of $N$ at the discontinuity $r^\star$.

In the present case we then considered that $N$ is well represented by 
\begin{equation}
N=\left\{
\begin{array}{lll}
 N_{\rm in}  &   {\rm for} & r < r_1^\star\\
 N_{\rm mid}  &   {\rm for} & r_1^\star < r < r_2^\star\\
 N_{\rm out}  & {\rm for} & r > r_2^\star \\
\end{array}
\right. ,
\label{brunt}
\end{equation}
with $N$ varying by the positive amounts $\Delta N_1= \left. N_{\rm in}\right|_{ r\rightarrow r^\star_{1-}}-\left. N_{\rm mid}\right|_ {r\rightarrow r^\star_{1+}}$ and $\Delta N_2= \left. N_{\rm out}\right|_{ r\rightarrow r^\star_{2+}}-\left. N_{\rm mid}\right|_ {r\rightarrow r^\star_{2-}}$ at $r=r_1^\star$  and $r=r_2^\star$, respectively.

Our starting point is the wave equation A1 of Appendix A of \cite{cunha19}.  In terms of the variable $\Psi=~(r^3/g\rho \tilde f) ^{1/2}\delta p$, where $\delta p$ is the Lagrangian pressure perturbation, $\rho$ is the density, $g$ is the gravitational acceleration, and $\tilde f$ is a function of frequency and of the equilibrium structure [the f-mode discriminant defined by equation 35 of \cite{gough07}], the equation resulting from the linear, adiabatic pulsation equations for the case of a spherically symmetric equilibrium
under the Cowling approximation, can be written as
\begin{eqnarray}
{\Psi^{\prime\prime}}+K^2\Psi=0,
\label{waveeqap1}
\end{eqnarray}
where a prime represents  a differentiation with respect to $r$. Given our focus on g modes, we approximate the radial wavenumber $K$ as 
 \begin{equation}
  K^2\approx -\frac{L^2}{r^2}\left(1-\frac{N^2}{\omega^2}\right),
\label{k2}   
 \end{equation} 
with $L^2=l(l+1)$ and $\omega$  the angular frequency of the mode.

The asymptotic solution satisfying the wave equation can be written in each region of the star as
\begin{equation}
\begin{array}{lcc}
 \Psi_{\rm in} \sim \tilde\Psi_{\rm in} K^{-1/2}
\sin\left(\int_{r_1}^r K\rmd r +
  \frac{\pi}{4}\right),  &  {\rm for} & r < r_1^\star   \\\Psi_{\rm mid} \sim \tilde\Psi_{\rm mid} K^{-1/2}
\sin\left(\int_r^{r_2} K\rmd r +
  \frac{\pi}{4}+\varphi\right),  &   {\rm for} & r_1^\star < r < r_2^\star\\
\Psi_{\rm out} \sim \tilde\Psi_{\rm out} K^{-1/2}  \sin\left(\int_r^{r_2} K\rmd r +
  \frac{\pi}{4}\right),  & {\rm for} & r > r_2^\star 
\end{array}
\end{equation}
where $\tilde\Psi_{\rm in}$, $\tilde\Psi_{\rm mid}$, and $\tilde\Psi_{\rm out}$ are constants and $\varphi$, whose functional form is to be derived, accounts for the phase shift induced by the glitch located at  $r=r_2^\star$.

Next we define the glitch amplitudes:\\

\noindent\as$_1=\left[(K_{\rm in}-K_{\rm mid})/K_{\rm mid}\right]_{r=r_1^\star}$\\ \as$_2=\left[(K_{\rm out}-K_{\rm mid})/K_{\rm mid}\right]_{r=r_2^\star}$.\\

Imposing the continuity of $\Psi$ and of its derivative at $r=r_1^\star$ we find the eigenvalue condition 
\begin{equation}
\sin\left(\int_{r_1}^{r_2}K\rmd 
  r+\frac{\pi}{2}+\varphi\right)=
-A_{\rm st}
\sin\left(\int_{r_1^\star}^{r_2}K\rmd
  r+\frac{\pi}{4}+\varphi\right)\cos\left(\int_{r_1}^{r_1^\star}K\rmd
  r+\frac{\pi}{4}\right).
\label{eq:eigen_step}
\end{equation}
We note that this condition is similar to the eigenvalue condition derived for mixed modes in the presence of a step-like glitch  in the inner half of the g-mode cavity \citep{cunha24}, but $\varphi$ is now to be interpreted as the phase shift induced by the outer glitch rather than the phase induced by the coupling with the p-mode cavity. Rewriting the first integral on right-hand-side of Eq.~(\ref{eq:eigen_step}) in terms of the other two integrals and developing the expression one can write the eigenvalue condition as
\begin{eqnarray}
\sin\left(\int_{r_1}^{r_2}K\rmd  r+\frac{\pi}{2}+\phi+\varphi\right) = 0,
\label{eigengap2}
\end{eqnarray}
where $\phi$ is the phase induced by the glitch located at $r=r_1^\star$ and is given by
\begin{equation}
\cot{\phi}=\frac{1+A_{\rm st1}\cos^2\left(\int_{r_1}^{r_1^\star} K_{\rm in}\rmd r+\frac{\pi}{4}\right)}{-A_{~\rm st1}\sin\left(\int_{r_1}^{r_1^\star} K_{\rm in}\rmd r+\frac{\pi}{4}\right)\cos{\left(\int_{r_1}^{r_1^\star} K_{\rm in}\rmd r+\frac{\pi}{4}\right)}}.
\end{equation}
The functional form of the phase $\varphi$ is derived by imposing the continuity of $\Psi$ and of its derivative at $r=r_2^\star$. This leads to the condition
\begin{equation}
\sin\varphi = -A_{\rm st2}\sin\left(\int_{r_2^\star}^{r_2} K_{\rm out}\rmd r+\frac{\pi}{4}+\varphi\right)\cos\left(\int_{r_2^\star}^{r_2} K_{\rm out}\rmd r+\frac{\pi}{4}\right).
\label{eq:eigen_r2}
\end{equation}
Developing the right-hand-side of Eq.~(\ref{eq:eigen_r2}) we then find that 
\begin{equation}
\cot{\varphi}=\frac{1+A_{\rm st2}\cos^2\left(\int_{r_2^\star}^{r_2} K_{\rm out}\rmd r+\frac{\pi}{4}\right)}{-A_{\rm st2}\sin\left(\int_{r_2^\star}^{r_2} K_{\rm out}\rmd r+\frac{\pi}{4}\right)\cos{\left(\int_{r_2^\star}^{r_2} K_{\rm out}\rmd r+\frac{\pi}{4}\right)}}.
\end{equation}
The fact that the two phases have the same functional form, which, in turn, is the same as that derived for a single step-like glitch in \cite{cunha19}, results from our option to model both glitches with a step function in the example given here.

The eigenvalue condition expressed by Eq.~(\ref{eigengap2}) differs from equation 5 of \cite{cunha24} only in the addition of the extra phase $\varphi$ introduced by the outer glitch. Moreover, it is mathematically equivalent to equation 36 of \cite{cunha15}, despite the different meanings of the phase $\varphi$. The derivations from Eq.~(\ref{eigengap2}) of the expressions for the reduced periods (Eq.~(\ref{eq:per_2g})) and reduced period spacings (Eq.~(\ref{eq:quadratic})) in the presence two glitches, thus follow the exact same steps as in these earlier works and we do not repeat them here. Noting that except near the turning points, $K\approx LN/r\omega$, we approximate the integrals in the definitions of $\phi$ and $\varphi$ so that $\int_{r_1}^{r^\star}K_{\rm in}\rmd  r +\pi/4 \approx \frac{\tilde{\mathcal{W}}_{\rm g}^\star}{2\pi} \Pi+\delta +\pi/4$ and $\int_{r_2^\star}^{r_2}K_{\rm out}\rmd  r +\pi/4 \approx \frac{\mathcal{W}_{\rm g}^\star}{2\pi} \Pi+\delta +\pi/4$, where $\delta$ accounts for the approximation made near the respective turning point. With this approximation, we recover the functional form for the step-like glitch $\phi$ and $\varphi$ in the form given by Eq~(\ref{eq:phase_step}).

\newpage
\section{Model fits}
\label{apC}
Corner plots and best-fit models for the selection of fits discussed in the main text. Note that in the corner plots the glitch positions are  expressed in terms of the degree-dependent buoyancy radius \wgstar\, and depth  \wgstart\, used in earlier works. 
To compare with the values provided for \wgstarlt\, in the tables in the main text, the values of \wgstart\, in the corner plots need to be divided by $\sqrt{2}$.

%\red{add tables with priors}

\begin{figure*}[!ht]
	\includegraphics[width=1.\columnwidth]{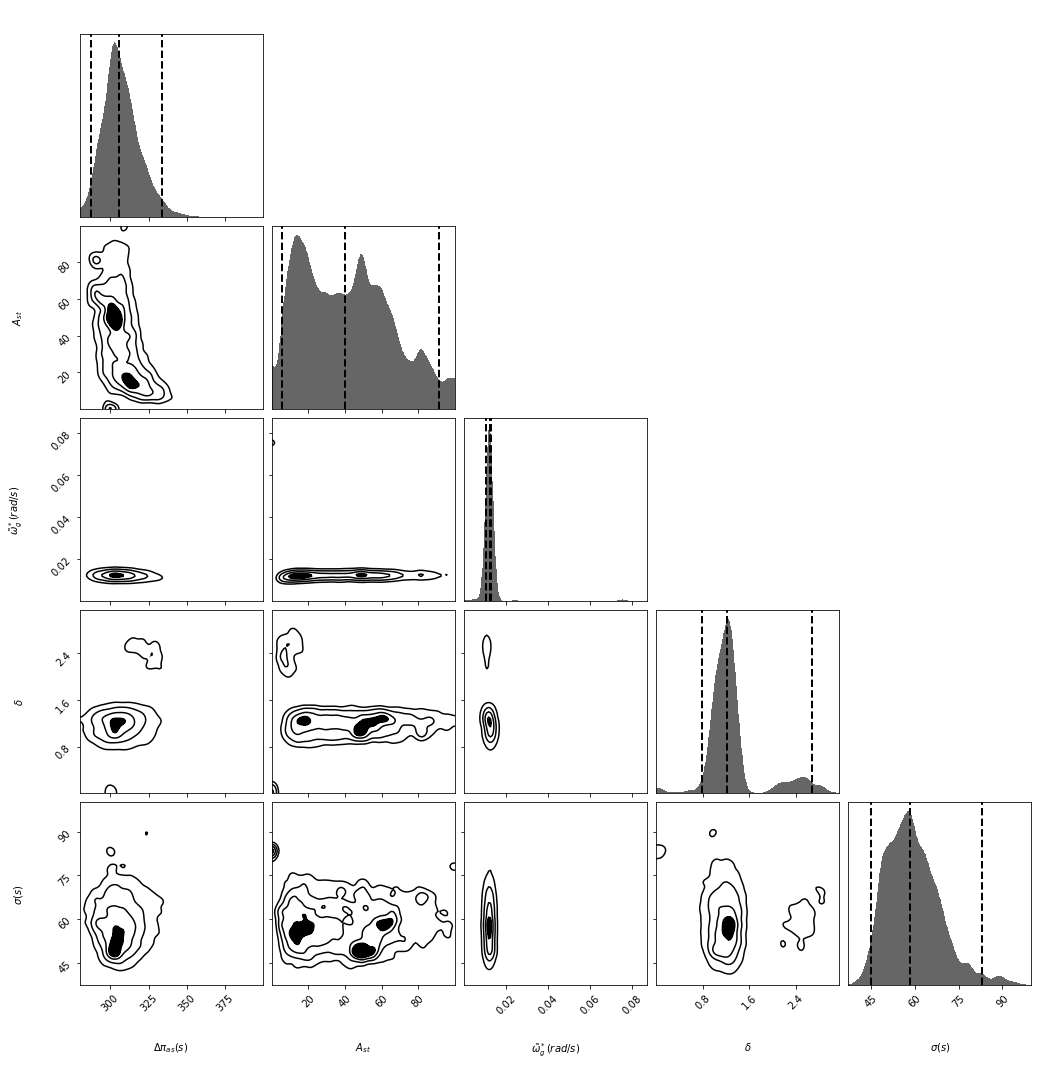}
    \caption{Posterior distributions for the model parameters resulting from the fit of the analytical model for one step-like glitch to the reduced period spacings of $l=1$ of KIC~10553698A, for reduced periods smaller than 11100~s (case A).}
        \label{fig:corner_KIC1_1}
\end{figure*}
\begin{figure*}
	\includegraphics[width=1.\columnwidth]{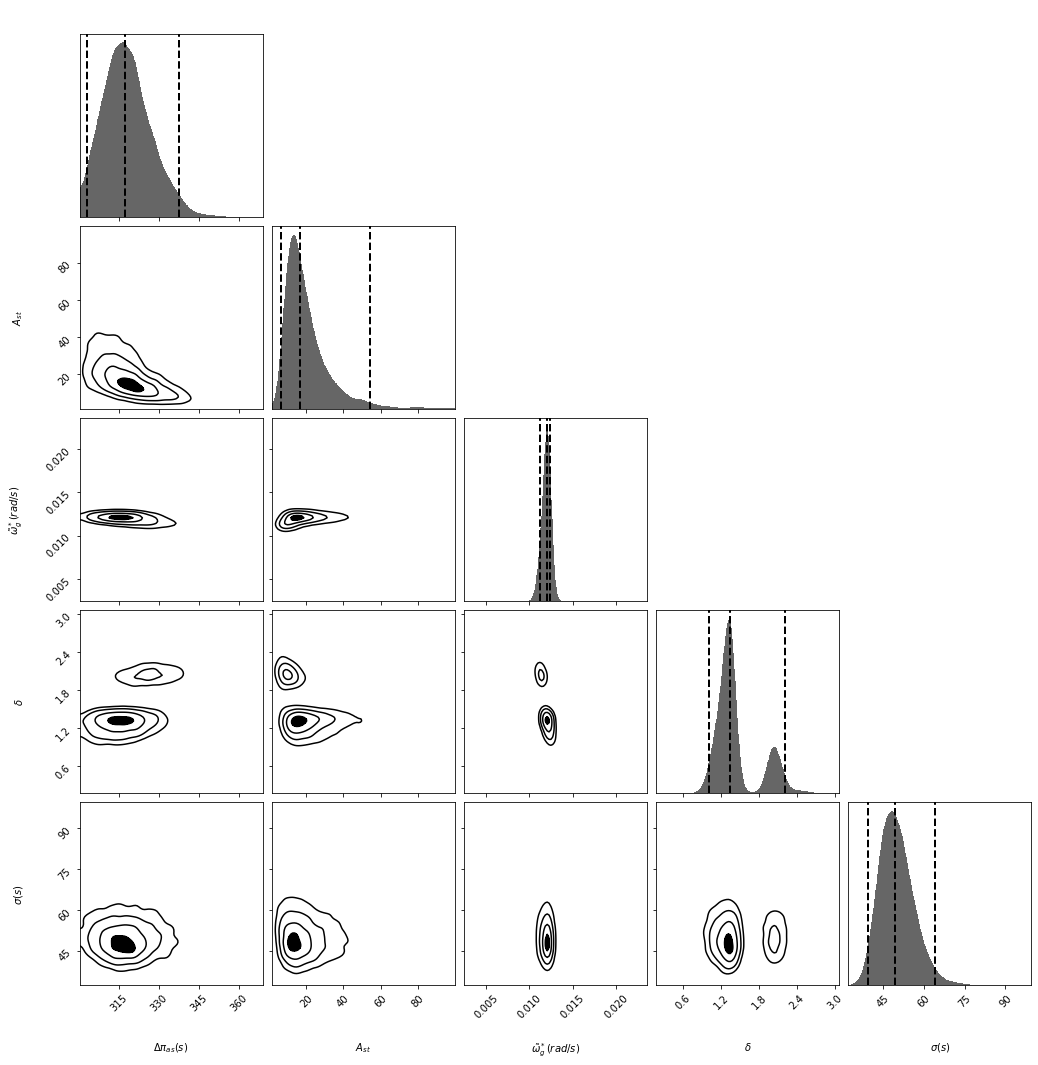}
    \caption{Posterior distributions for the model parameters resulting from the fit of the analytical model for one step-like glitch to the reduced period spacings of $l=1$ and $l=2$ of KIC~10553698A, for reduced periods smaller than 11100~s (case B).  }
        \label{fig:corner_KIC1_2}
\end{figure*}

\begin{figure*}
	\includegraphics[width=1.\columnwidth]{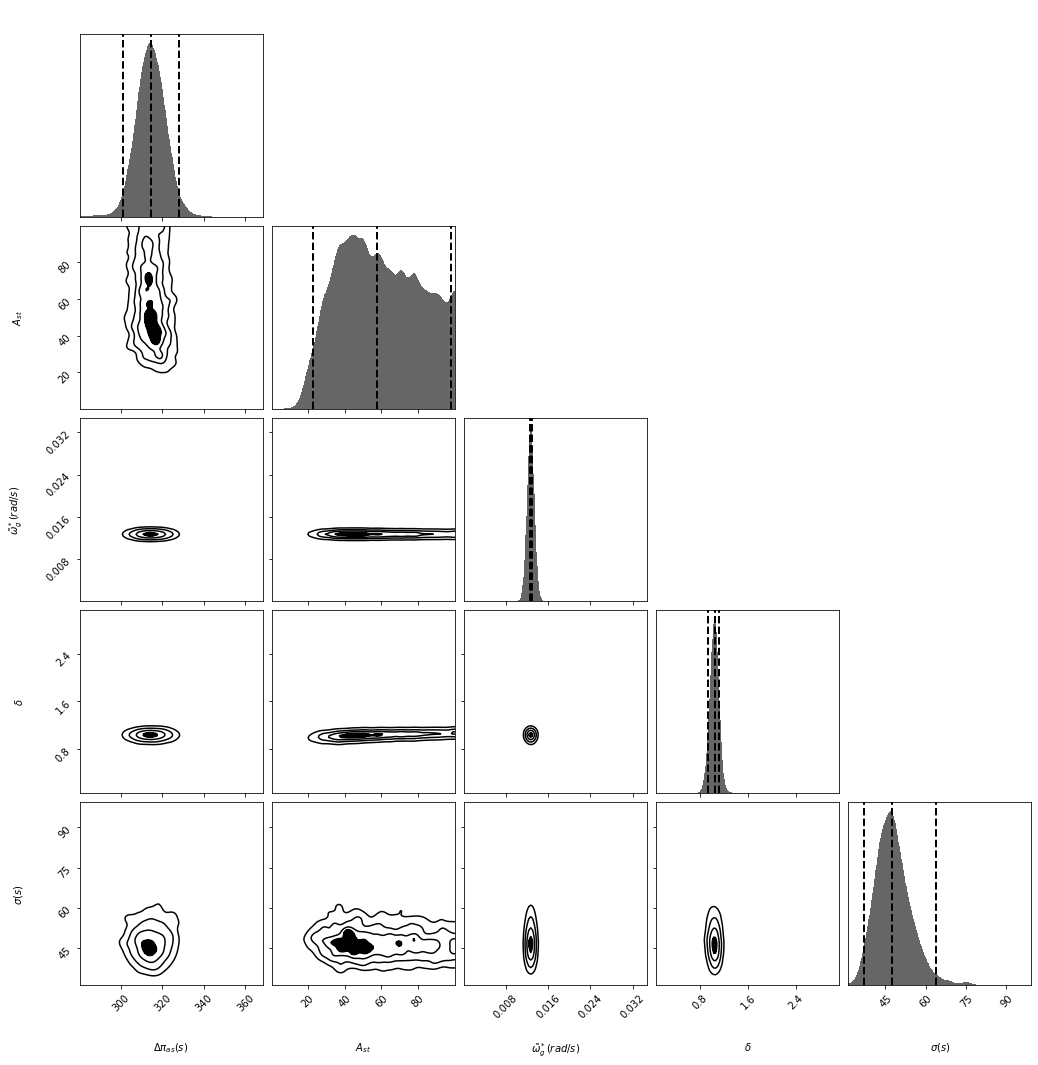}
    \caption{Posterior distributions for the model parameters resulting from the fit of the analytical model for one step-like glitch to the reduced period spacings of $l=1$ of EPIC~211779126 (case F).    
}
        \label{fig:corner_EPIC_1} 
\end{figure*}
\begin{figure*}
	\includegraphics[width=1.\columnwidth]{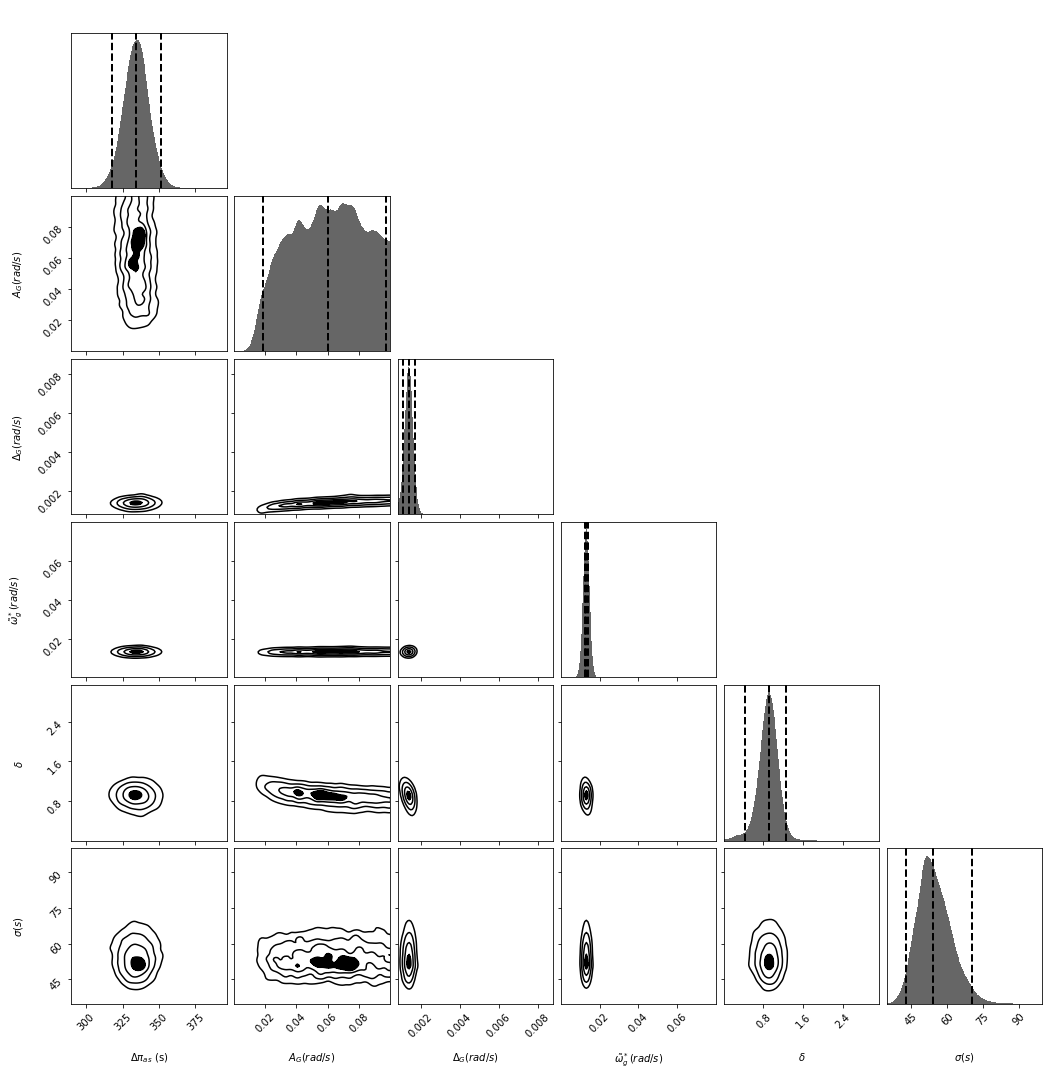}
    \caption{Posterior distributions for the model parameters resulting from the fit of the analytical model for one Gaussian-like glitch to the reduced period spacings of $l=1$ of EPIC~211779126 (case H). 
}
                \label{fig:corner_EPIC_2} 
\end{figure*}

\begin{figure*}
	\includegraphics[width=1.\columnwidth]{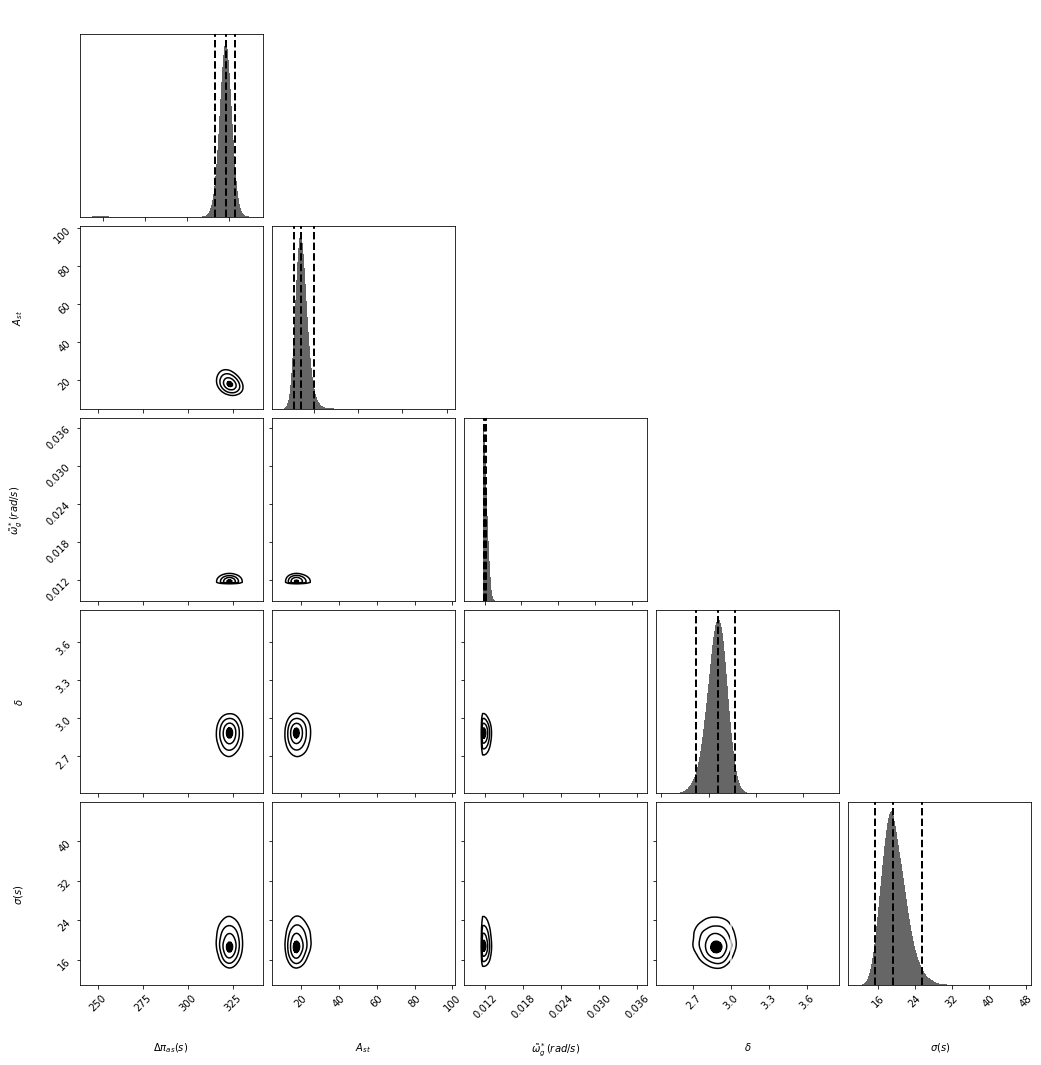}
    \caption{Posterior distributions for the model parameters resulting from the fit of the analytical model for one step-like glitch to the reduced period spacings computed from the $l=1$ and $l=2$ modes of KIC~10001893 in the region of periods where data on both mode degrees are available (case J).  
}
       \label{fig:corner_KIC2_1}
\end{figure*}
\FloatBarrier
\begin{figure*}[!ht]
	\includegraphics[width=1.\columnwidth]{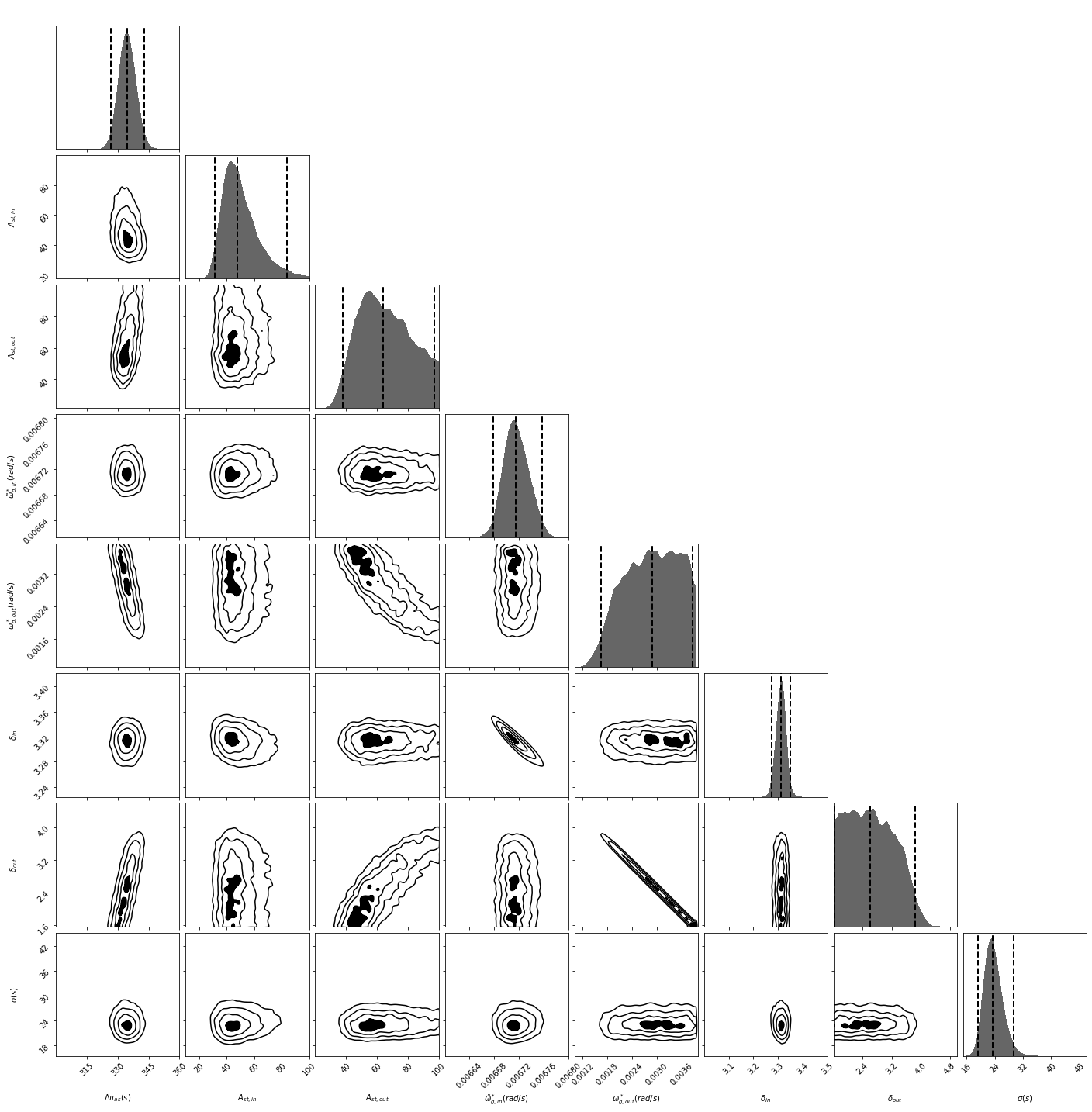}
    \caption{Posterior distributions for the model parameters resulting from the fit of the analytical model for two step-like glitches to all reduced period spacings computed from the $l=1$ and $l=2$ modes of KIC~10001893 (case K). 
}
        \label{fig:corner_KIC2_2} 
\end{figure*}

\begin{table}[!ht]
\centering
\caption{Best model parameters for cases A and B in Table~\ref{tab:KIC1_parameters}.}
\begin{tabular}{lcc}
\toprule
\textbf{Parameter} & \textbf{Case A} & \textbf{Case B} \\
\midrule
$\Delta \pi_{\rm as}$ (s)         & 326.10 & 317.07 \\
$A_{\rm st}$                      & 7.93   & 17.15  \\
$\tilde{\omega}_{\rm g/L}^*$ (rad/s) & 0.007928 & 0.008498 \\
$\delta$                      & 2.183  & 1.345  \\
\bottomrule
\end{tabular}
\end{table}

\begin{table}[ht]
\centering
\caption{Best model parameters for cases C and E in Table~\ref{tab:KIC1_parameters}.}
\begin{tabular}{lcc}
\toprule
\textbf{Parameter} & \textbf{Case C} & \textbf{Case E} \\
\midrule
$\Pi_{\rm s,\min}$ (s)            & 3615.23 & 3587.59 \\
$\Delta \pi_{\rm as}$ (s)         & 325.42  & 328.52  \\
$A_{\rm st}$                      & 38.57   & 17.82   \\
$\tilde{\omega}_{\rm g/L}^*$ (rad/s) & 0.007874 & 0.007094 \\
$\delta$                      & 0.687   & 1.604   \\
\bottomrule
\end{tabular}
\end{table}

\begin{table}[ht]
\centering
\caption{Best model parameters for cases A and  B in Table~\ref{tab:EPIC_parameters}.}
\begin{tabular}{lcc}
\toprule
\textbf{Parameter} & \textbf{Case A} & \textbf{Case B} \\
\midrule
$\Delta \pi_{\rm as}$ (s)         & 313.27 & 316.73 \\
$A_{\rm st}$                      & 96.47   & 79.68 \\
$\tilde{\omega}_{\rm g/L}^*$ (rad/s) &  0.008974 & 0.007865 \\
$\delta$                      & 1.063  & 2.381 \\
\bottomrule
\end{tabular}
\end{table}

\begin{table}[ht]
\centering
\caption{Best model parameters for case C  and D in Table~\ref{tab:EPIC_parameters}.}
\begin{tabular}{lcc}
\toprule
\textbf{Parameter} & \textbf{Case C} & \textbf{Case D}\\
\midrule
$\Delta \pi_{\rm as}$ (s)         & 335.00  & 338.54\\
$A_{\rm G}$ (rad/s)                      & 0.03892  & 0.04633  \\
$\tilde{\omega}_{\rm g/L}^*$ (rad/s) & 0.009256 & 0.008128 \\
$\Delta_{\rm g}$ (rad/s)            & 0.001294 & 0.002069\\
$\delta$                      & 0.975  & 2.330 \\
\bottomrule
\end{tabular}
\end{table}

\begin{table}[ht]
\centering
\caption{Best model parameters for cases A and  B in Table~\ref{tab:KIC2_parameters}.}
\begin{tabular}{lcc}
\toprule
\textbf{Parameter} & \textbf{Case A} & \textbf{Case B} \\
\midrule
$\Delta \pi_{\rm as}$ (s)          & 323.90 & 335.75\\
$A_{\rm st}$                         & 17.12 & 43.00\\
$\tilde{\omega}_{\rm g/L}^*$ (rad/s) & 0.008507 & 0.004741\\
$\delta$                       & 2.89 & 3.322 \\
\hline
$A_{\rm st}$                         & -- &  58.53\\
$\tilde{\omega}_{\rm g/L}^*$ (rad/s)  & -- & 0.001986\\
$\delta$                       & -- & 2.688 \\
\bottomrule
\end{tabular}
\tablefoot{ The bottom part of the table shows the results for the second glitch in the two-glitch model adopted in case B.}
\end{table}

%%%%%%%%%%%%%%%%%%%%%%%%%%%%%%%%%%%%%%%%%%%%%%%%%%

\end{appendix}
% Don't change these lines
%\bsp	% typesetting comment
%\label{lastpage}
\end{document}